\documentclass[twocolumn,showpacs,amsmath,prd,amssymb,showkeys,superscriptaddress]{revtex4}

\pdfoutput=1

\usepackage{graphicx}
\usepackage{dcolumn}
\usepackage{multirow}
\usepackage{mathtools}
\usepackage{scalerel}
\usepackage{bm}

\def\0nubb{0 \nu \beta \beta}
\def\2nubb{2 \nu \beta \beta}
\def\tauhalf{T_{\scaleto{1/2}{5pt}}}
\def\thalf0nu{T^{0 \nu}_{\scaleto{1/2}{5pt}}}

\def\thl2{T^{2 \nu}_{\scaleto{1/2}{5pt}}}
\def\mbb{ \langle m_{\beta \beta} \rangle}
\def\mbbmin{ \langle m_{\beta \beta} \rangle _{-}}
\def\mbbmax{ \langle m_{\beta \beta} \rangle _{+}}
\def\p95mbb{ \langle m_{\beta \beta} \rangle _{95\%}}
\def\Qbb{ Q_{\beta \beta} }
\def\Nbb{ A_{\beta \beta} }
\def\Abb{ A_{\beta \beta} }
\def\s0nu{ S_{0 \nu} }
\def\b2nu{ B_{2 \nu} }
\def\f2nu{ f_{2 \nu} }

\def\P503s{{\rm P_{50}^{3\sigma}}}
\def\ME2{| M^{0 \nu} | ^2}

\def\FWHM{w_{\scaleto{1/2}{4pt}}}
\def\FWFM{w_{3 \sigma}}
\def\tyr{{\rm ton\mbox{-}yr}}

\def\BIunit{{\rm counts}/( \FWHM \mbox{-} \tyr )} 
\def\Bnowunit{{\rm count}/( \FWHM \mbox{-} \tyr )} 
\def\BM{{\rm count}/( \FWHM \mbox{-} \Sigma )} 
\def\kty{/( {\rm keV} \mbox{-} \tyr )}

\def\ca48{\rm ^{48}Ca}
\def\ge76{\rm ^{76}Ge}
\def\se82{\rm ^{82}Se} 
\def\zr96{\rm ^{96}Zr}
\def\mo100{\rm ^{100}Mo}
\def\cd116{\rm ^{116}Cd}
\def\te130{\rm ^{130}Te}
\def\xe136{\rm ^{136}Xe}
\def\nd150{\rm ^{150}Nd}

\def\B0{{\rm B_0}}
\def\S0{{\rm S_0}}
\def\BIcurrent{{\rm BI_0}}
\def\BI0{{\rm BI}}
\def\Sref{{\rm S_{ref}}}

\begin{document}


\title{
The Exposure-Background Duality 
in the Searches of
Neutrinoless Double Beta Decay
}

%

\newcommand{\as}{Institute of Physics, Academia Sinica,
Taipei 11529}
\newcommand{\bhu}{Department of Physics, Institute of Science,
Banaras Hindu University,
Varanasi 221005}
\newcommand{\cusb}{Department of Physics, 
School of Physical and Chemical Sciences, 
Central University of South Bihar, Gaya 824236}
\newcommand{\thu}{Department of Engineering Physics, 
Tsinghua University, Beijing 100084}
\newcommand{\corr}{htwong@phys.sinica.edu.tw}

\author{ M.K.~Singh } \affiliation{ \as } \affiliation{ \bhu }
\author{ H.T.~Wong } \altaffiliation[Corresponding Author: ]{ \corr } \affiliation{ \as }
\author{ L.~Singh }  \affiliation{ \as } \affiliation{ \cusb }
\author{ V.~Sharma }  \affiliation{ \as } \affiliation{ \bhu }
\author{ V.~Singh }  \affiliation{ \as } \affiliation{ \bhu } \affiliation{ \cusb }
\author{ Q.~Yue }  \affiliation{ \thu } 




\date{\today}

\begin{abstract}

Tremendous efforts are required
to scale the summit of observing neutrinoless
double beta decay ($\0nubb$).  This article quantitatively
explores the interplay between exposure 
(target mass$\times$data taking time ) 
and background levels in $\0nubb$ experiments.
In particular, background reduction can
substantially alleviate the necessity of 
unrealistic large exposure as the normal
mass hierarchy (NH) is probed.
The non-degenerate (ND)-NH can be covered with an exposure 
of $\mathcal{O}$(100)~ton-year, 
which is only an order of magnitude 
larger than those planned for
next generation projects $-$ provided that the background 
could be reduced by $\mathcal{O}(10^{-6})$
relative to the current best levels.
It follows that background suppression will be playing
increasingly important and investment-effective, 
if not determining, roles in 
future $\0nubb$ experiments 
with sensitivity goals of 
approaching and covering ND-NH.

\end{abstract}

\pacs{
14.60.Pq,
23.40.-s,
02.50.-r.
}
\keywords{
Neutrino Mass and Mixing,
Double Beta Decay,
Statistics. 
}

\maketitle


\section{Introduction}

The nature of the neutrinos~\cite{nu-review}, 
and in particular
whether they are Majorana or Dirac particles, 
is an important problem in particle physics,
the answer to which will have profound implications
to the searches and formulation of physics beyond
Standard Model and the Grand Unified Theories.
Neutrinoless double beta decay ($\0nubb$) 
is the most sensitive experimental probe 
to address this question~\cite{nubb-review}. 
Observation of $\0nubb$ implies: 
(i) that neutrinos are Majorana particles, and 
(ii) lepton number violation.
Since several decades, there are intense activities world-wide
committed to the experimental searches of $\0nubb$. 
 
Neutrino oscillation experiments~\cite{nu-review,nuosc-review}
are producing increasingly precise information
on the mass differences and mixings among the three
neutrino mass eigenstates.
The latest data imply slight preferences
of the ``Normal Hierarchy'' (NH) over
the ``Inverted Hierarchy'' (IH) in
the structures of the neutrino mass eigenstates~\cite{NHoverIH}.
In parallel,
cosmology data~\cite{nucosmo-review}
provide stringent upper bounds
on the total mass of the neutrinos,
with good prospects on an actual measurement 
in the future.
Together, a picture emerges providing a glimpse 
on the parameter space where
positive observations of $\0nubb$  may reside.
Experimental studies are expected to require significant efforts 
and resources
$-$  especially so if NH is confirmed.
Detailed quantitative studies on the optimal strategies
``to scale this summit'' with finite resources
would be highly necessary.

The current work addresses one aspect
of this issue.
We studied the required exposures of 
$\0nubb$-projects
versus the expected background $\B0$ 
before the experiments are performed.
The notations and formulation are described 
in Section~\ref{sect::formulation}.
The effects on the ``discovery potentials''
with varying $\B0$, and the
implied experimental strategies,
are discussed in Section~\ref{sect::strategy}.
The connections with the current
landscape in neutrino physics are made
in Section~\ref{sect::nuphys} via the choice
of a particular model on the evaluation of nuclear
matrix elements.
Various aspects on the interplay between
exposure and background in $\0nubb$ experiments 
are discussed in Section~\ref{sect::duality}

Background typically includes two generic components
each having different energy dependence $-$
the ambient background and the irreducible
intrinsic background from 
cosmogenic radioactivity and 
two-neutrino double beta decay ($\2nubb$).
Only the combined background is considered 
in this work, while on-going research
efforts are attending the different roles of 
the two components.
In particular, the constraints imposed by the $\2nubb$ background
to detector resolution are discussed in
Section~\ref{sect::2nubkg}.

\section{Formulations and Notations}
\label{sect::formulation}

\subsection{Double Beta Decay}
\label{sect::dbd}

The process $\0nubb$ in candidate
nucleus $\Nbb$
refers to the decay
\begin{equation}
^N_Z \Abb ~ \rightarrow ~
_{Z+2}^{N-2}A ~ + ~ 2 e^-  ~~.
\label{eq::0nubb}
\end{equation}
The experimental signature
is distinctive.
The summed kinetic energy
of the two emitted electrons
corresponds to a peak at the
transition Q-value ($\Qbb$),
which is known and unique for
each $\Nbb$.

The width of the $\0nubb$-peak
(denoted by $\Delta$ in \%) 
characterizes the energy resolution of the detector,
and is defined $-$ a natural choice and also
following convention in the literature 
for clarity $-$ 
as the ratio of 
full-width-half-maximum (FWHM, denoted by $\FWHM$) to
the total measureable energy $\Qbb$,
such that $ \FWHM {=} ( \Delta {\cdot} \Qbb )$.

Various beyond-standard-model processes
invoking lepton-number-violation
can give rise to $\0nubb$~\cite{nubb-bsm}.
In the case of the ``mass mechanism''
where $\0nubb$ is driven by the Majorana neutrino mass,
the $\0nubb$ half-life ($\thalf0nu$)
can be expressed by~\cite{nubb-review,nubb-Robertson}
\begin{equation}
\left[ \frac{1}{ \thalf0nu } \right] ~ = ~
G^{0 \nu} ~ g_A^4 ~ \ME2 ~ \left| \frac{\mbb}{m_e} \right| ^2
\label{eq::thalf0nu}
\end{equation}
where
$m_e$ is the electron mass,
$g_A$ is the effective axial vector coupling~\cite{nubb-gA}, 
$G^{0 \nu}$ is a known phase space
factor~\cite{phasespace} due to kinematics,
$| M^{0 \nu} |$ is the nuclear physics
matrix element~\cite{dbdme-review}, while
$\mbb$ is the effective Majorana neutrino mass term
which depends on 
neutrino masses ($m_{i}$ for eigenstate $\nu_i$) 
and mixings ($U_{ei}$ for the component of $\nu_i$ in $\nu_e$):
\begin{equation}
\label{eq::mbb}
\mbb ~ = ~ 
| ~ U_{e1}^2 ~  m_{1} ~ + 
~ U_{e2}^2  ~ m_{2}  ~ e^{i \alpha} ~ +
~ U_{e3}^2  ~ m_{3}  ~ e^{i \beta} ~
| ~  
\end{equation}
where $\alpha$ and $\beta$ are the Majorana phases. 


The measureable half-life $\thalf0nu$
from an experiment which observes
$N_{obs}^{0 \nu}$-counts of $\0nubb$-events
in time $t_{\rm DAQ}$
in a ``Region-of-Interest'' (RoI)
at an efficiency of $\varepsilon_{RoI}$ 
can be expressed as 
\begin{equation}
{  \thalf0nu }  ~ = ~
 { {\rm ln~2} }   \cdot 
 N ( \Abb  )   \cdot t_{\rm DAQ} \cdot
\left[ \frac{\varepsilon_{RoI}}{N_{obs}^{0 \nu}} \right]  
\label{eq::natom}
\end{equation}
where $N ( \Abb )$ is the number of $\Abb$ atoms 
being probed.

For simplicity in discussions and to allow
the results be easily convertible to 
different configurations $-$
while capturing the essence of the physics, 
results in this article are derived in
the special ``ideal'' case where the target is
made up of completely enriched $\Abb$ isotopes.
That is, the isotopic abundance~(IA) is 100\%.
In additional, the various experimental efficiency
factors are all unity ($\varepsilon_{expt} {=} 100\%$).
Accordingly, Eq.~\ref{eq::natom} becomes
\begin{equation}
{  \thalf0nu }  ~ = ~
 { {\rm ln~2} }   \cdot 
\left[ \frac{N_A}{M(\Abb)} \right] \cdot \Sigma \cdot
\left[ \frac{\varepsilon_{RoI}}{N_{obs}^{0 \nu}} \right]  
\label{eq::roi}
\end{equation}
where $N_A$ is the Avogadro Number, 
$M ( \Abb )$ is the molar mass of $\Abb$, 
and $\Sigma$ denotes
the combined exposure (mass$\times$$t_{\rm DAQ}$)
expressed in units of ton-year~(ton-yr)
at $\Abb$ at IA=100\% and $\varepsilon_{expt} {=} 100\%$.
Effects due to these parameter choices and other assumptions
will be discussed in Section~\ref{sect::choice} where
conversion relations to those for realistic experiments
are given.

The expression of Eq.~\ref{eq::roi} applies to experiments
with counting analysis. More sophisticated statistical methods
are usually adopted to extract full information from a given
data set. These typically exploit the energy spectral shapes,
which are known for the signal and 
are predictable with uncertainties for the background.
However, in the conceptual-design and 
sensitivity-projection stage of experiments, 
the simplified and intuitive approach of
Eq.~\ref{eq::roi} will suffice,
especially so in the low count rate Poisson statistics
regime which is of particular interest in this article.

Combining the theoretical and experimental descriptions
of $\thalf0nu$ from, respectively, 
Eqs.~\ref{eq::thalf0nu}\&\ref{eq::roi} gives:
\begin{eqnarray}
\ME2  \left[ g_A^4 \cdot  H^{0 \nu} \right] & = & 
 \frac{1}{\mbb ^2} 
\left[ \frac{1}{\Sigma} \cdot
\frac{N_{obs}^{0 \nu}}{\varepsilon_{RoI}} \right] ~~ , 
~~ {\rm where} \nonumber \\
H^{0 \nu}  & \equiv & 
{\rm ln~2} ~ \left[ \frac{N_A}{M ( \Abb ) \cdot m_e^2} \right] ~ G^{0 \nu} ~~
\label{eq::H0nu}
\end{eqnarray}
is called ``specific phase space'' 
in the literature~\cite{nubb-Robertson}.


\subsection{Discovery Potential}
\label{sect::strategy}

In our context, $\B0$ is
expected background counts within the RoI around $\Qbb$.
This can, in principle, be predicted with good accuracies
prior to the experiments.
The sensitivity goals of experiments
are typically expressed in the literature~\cite{mbbprob} as:
``Discovery Potential at 3$\sigma$ with 50\% probability''
($\P503s$) and
``upper limits at 90\% confidence level''
which characterize possible positive and negative outcomes, respectively.
We focus on $\P503s$ in this work, for the reason
that next-generation $\0nubb$ experiments should be designed
to have the maximum reach of discovery, rather than setting
limits.


\begin{figure}
{\bf (a)}\\
\includegraphics[width=8.2cm]{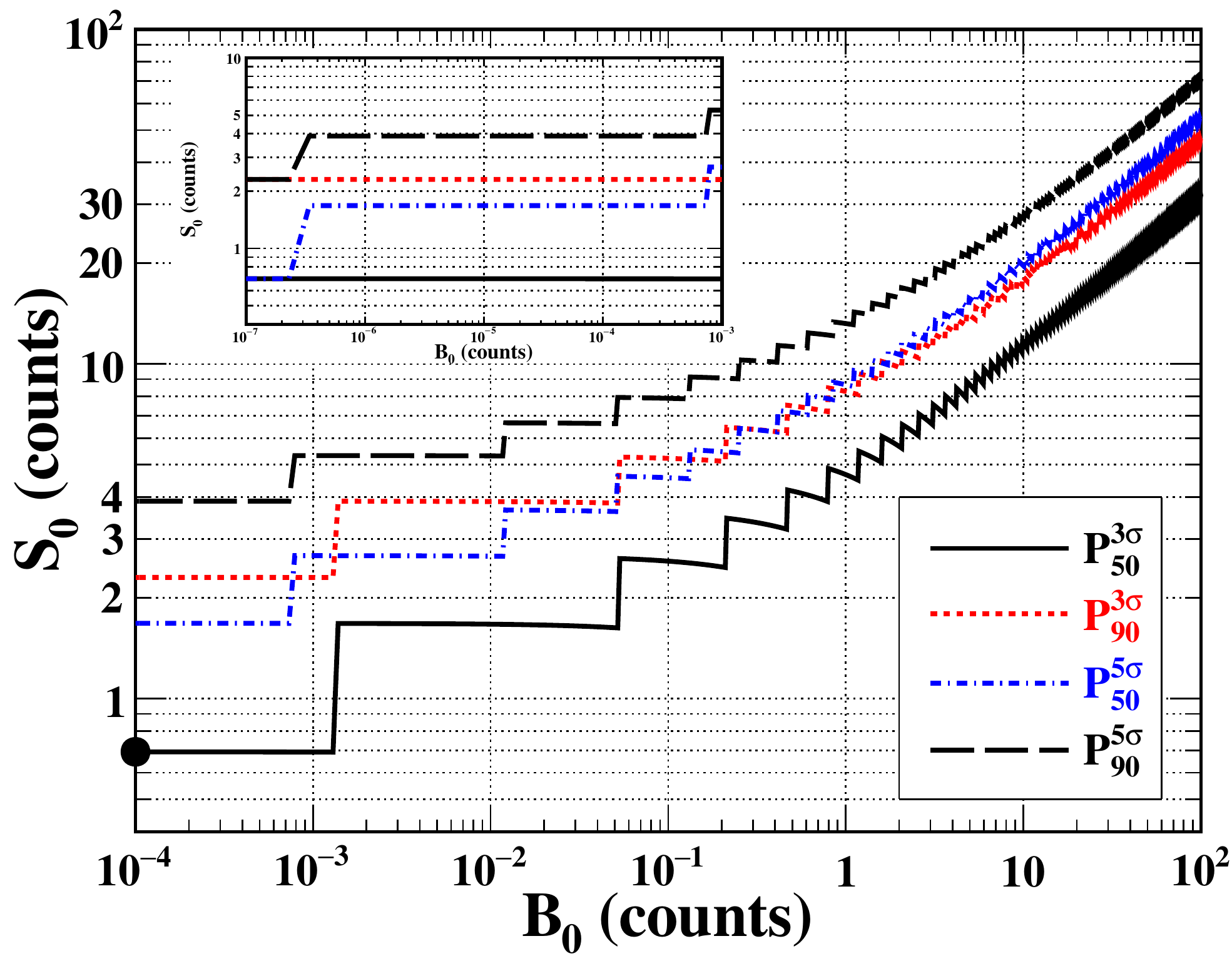}\\
{\bf (b)}\\
\includegraphics[width=8.2cm]{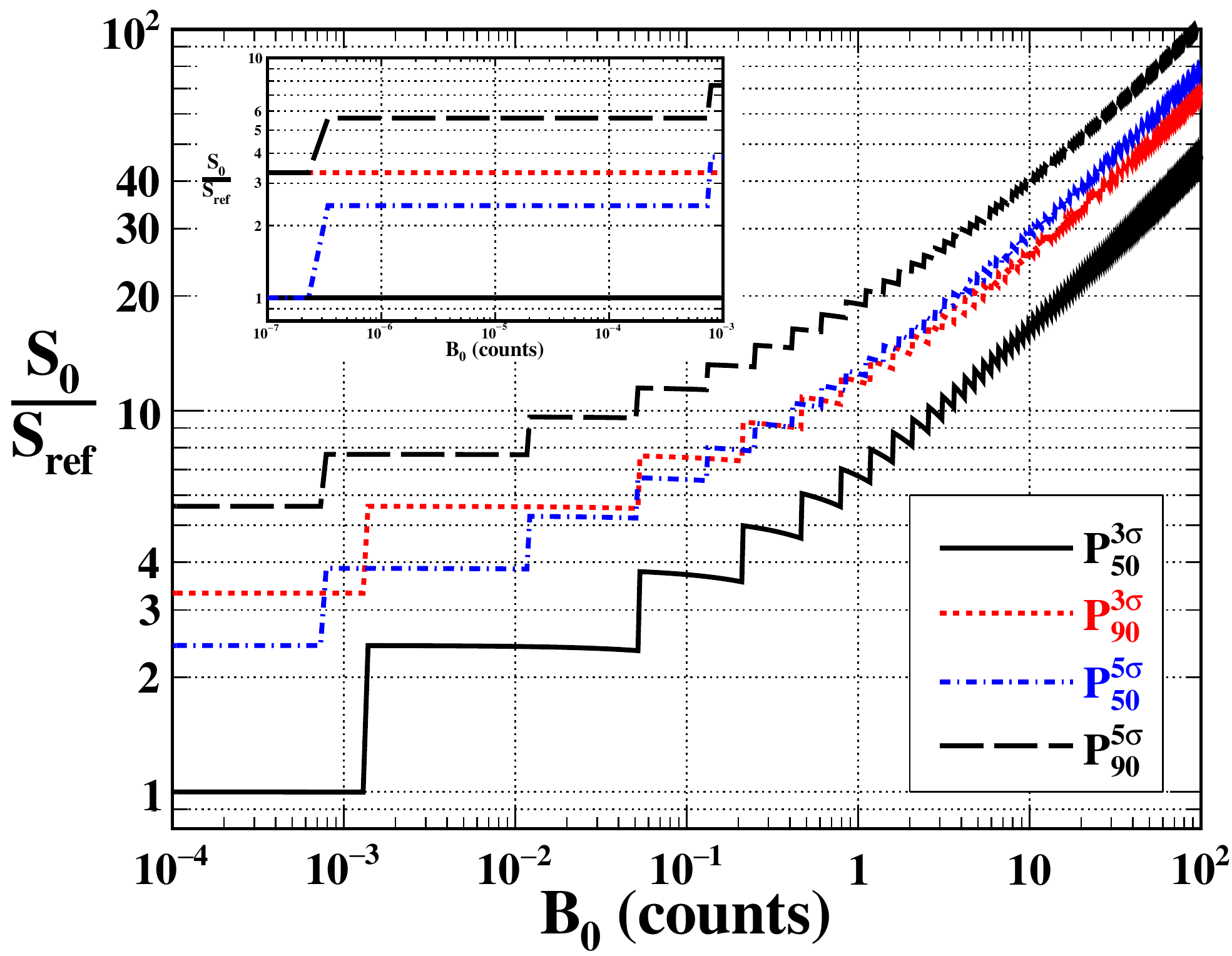}
\caption{
The variations of 
(a) $\S0$ 
and 
(b) ratios of $\S0$ to $\Sref$
versus $\B0$ in the
discovery potential of $\geq 3 \sigma , 5 \sigma$ with $\geq 50\% , 90\%$.
The specific case of $\P503s$ at
3$\sigma$ and 50\% is adopted as the criteria in this work.
The level of $\Sref$ at the background-free condition for $\P503s$ 
is represented with a black dot in (a)
}
\label{fig::S0vsB0}
\end{figure}


Poisson statistics is necessary to handle
low background and rare signal processes.
The dependence of 
the required average signal ($\S0$) versus $\B0$  
under $\P503s$ 
and other discovery potential criteria
are depicted in Figure~\ref{fig::S0vsB0}a.
For a given real and positive $\B0$ as input and
using $\P503s$ as illustration,
the Poisson distribution $P(i;\mu)$ 
is constructed with mean $\mu {=} \B0$.
The observed count $N_{obs}^{3 \sigma}$
is evaluated as the smallest integer which
satisfies  
\begin{equation}
\sum_{i=0}^{N_{obs}^{3 \sigma}} P ( i ; \B0 ) \geq ( 1 - 0.00135 )
\label{eq::B0}
\end{equation}
where 0.00135 is the fraction of a
Gaussian distribution in the interval
$[ {+} 3 \sigma , \infty ]$. 
This is the minimal observed event integer number with
${\geq} 3 \sigma$ significance over
a predicted average background $\B0$. 
The output $\S0$ is the minimal signal strength
corresponding to the case where
the average total event  
$( \B0 {+} \S0 ) {=} N_{obs}^{3 \sigma} $ with
$\geq$50\% probability.
This is evaluated as the
minimum value which satisfies 
another Poisson distribution 
under the condition:
\begin{equation}
\sum_{i=0}^{N_{obs}^{3 \sigma}} P ( i ; [ \B0 {+} \S0 ] ) \geq  0.5 ~~.
\label{eq::B0S0}
\end{equation}

It can be inferred from Figure~\ref{fig::S0vsB0}a
that the ``background-free'' level 
with $\P503s$ criteria corresponds to a background
of $\B0 {<} 10^{-3}$ and a reference-point
signal of $\S0 {\equiv} \Sref {=} 0.69$. 

The ratios of $\S0$ relative to $\Sref$
are depicted in Figure~\ref{fig::S0vsB0}b.
It can be seen that one would require 
factors of 6.7(16.8) stronger signals 
to establish positive results at $\P503s$ 
when $\B0$ increases from $< 10^{-3}$ to  1(10).

While the predicted average background $\B0$ can be continuous and real numbers,
only integer counts can be observed in an experiment.
This gives rise to the relations being inequalities
in Eqs.~\ref{eq::B0}\&\ref{eq::B0S0}
and consequently the steps in Figures~\ref{fig::S0vsB0}a\&b.
In addition, signal and background events are indistinguishable experimentally.
The $\P503s$ criteria is applied to $(\B0 {+} \S0 )$ 
versus $\B0$, while the $\S0$ dependence on $\B0$ is
shown in Figures~\ref{fig::S0vsB0}a\&b.
This is the origin of the negative slopes in various segments.

\subsection{Background Index}

The theme of this work is to study the
interplay between required exposure and background
in $\0nubb$ experiments to meet certain
$\mbb$ target sensitivities.

In realistic experiments,
it is more instructive to characterize
background with respect to exposure
and the RoI energy range, such that
the relevant parameter 
is the 
``Background Index'' ($\BI0$)
defined as:
\begin{equation}
\BI0 ~  \equiv  ~ 
 \frac{\B0 ( {\rm RoI} ) }{ \Sigma }  
\label{eq::BI}
\end{equation}
which is the background within the RoI 
(chosen to be $\equiv \FWHM$, following convention)
per 1~ton-year of exposure,  
with dimension [$\BIunit$]. 
Background levels expressed in $\BI0$ 
are universally applicable to 
compare sensitivities of varying $\Abb$ 
in different experiments.


\subsection{Conversion to Realistic Configurations}
\label{sect::choice}

As explained in Section~\ref{sect::dbd},
the  $( \BI0 , \Sigma )$ results presented in this article
correspond to the ideal case where
IA=100\% and $\varepsilon_{expt} {=} 100\%$.
In addition, while 
the range of $g_A {\in} [0.6 , 1.27]$ 
is generally considered possible~\cite{nubb-Robertson,nubb-gA},
the ``unquenched'' free nucleon value of $g_A {=} 1.27$ 
is adopted.

\begin{figure}
\includegraphics[width=8.2cm]{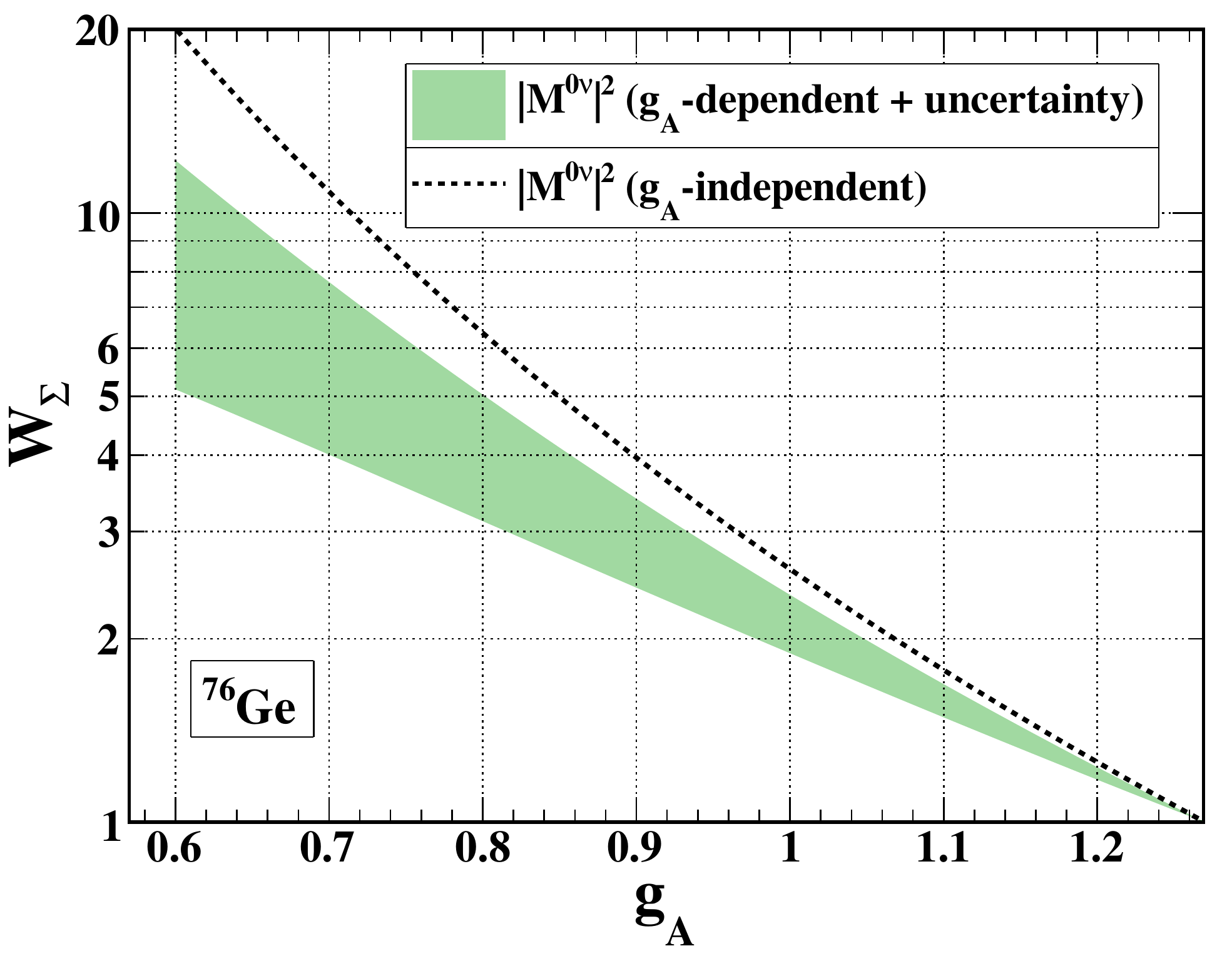}
\caption{
Variation of $W_\Sigma$ as defined in Eq.~\ref{eq::scaling} due
to changes in $g_A$ relative to that of $g_A$=1.27 in the case of $^{76}$Ge. 
The finite band width is the consequence of
the spread in $\ME2$ predictions~\cite{dbdme-review,me-fctga}.
The specific case where $\ME2$ is independent of $g_A$
such that $\Sigma {\propto}  [ g_A ] ^{\mbox{-}4} $
is denoted by the dotted line.
}
\label{fig::gAdependence}
\end{figure}

The required exposure $( \Sigma ' )$ in realistic experiments
would be larger and 
can be readily converted from the $\Sigma$-values via
\begin{equation}
 \Sigma '  ~ \simeq  ~  \Sigma
\cdot 
\frac{ 1 }{  {\rm IA} } 
\cdot 
\frac{ 1 }{ \varepsilon_{expt} }
\cdot 
W_{\Sigma} (g_A) ~~ ,
\label{eq::scaling}
\end{equation}
where $W_{\Sigma} (g_A)$ is the weight factor for $\Sigma$ due to the 
$g_A$-dependence~\cite{dbdme-review,me-fctga} 
of $\thalf0nu$ in Eq.~\ref{eq::thalf0nu},
relative to the values at $g_A$=1.27. 
It is depicted in Figure~\ref{fig::gAdependence} 
for the case of $\ge76$.
The finite band width as a function of $g_A$
is the consequence of
the spread in $\ME2$ predictions~\cite{dbdme-review,me-fctga}.
The specific case where $\ME2$ is independent of $g_A$
implies $\Sigma {\propto} [ g_A ]^{\mbox{-}4}$ 
and is denoted by the dotted line.

The background index defined relative to $\Sigma '$ 
for realistic configurations
can accordingly be expressed as
\begin{equation}
 \BI0 ' ( \Sigma ' )  ~ \simeq  ~  \BI0  \cdot 
\left[ \frac{\Sigma}{\Sigma '} \right]  ~~~ ,
\end{equation}
such that  $\Sigma ' {>} \Sigma$ and $\BI0 ' {<} \BI0$.
Realistic experiments naturally imply larger exposure and more 
stringent background requirements.


\subsection{Neutrino Physics Connections}
\label{sect::nuphys}


\begin{figure}
\includegraphics[width=8.2cm]{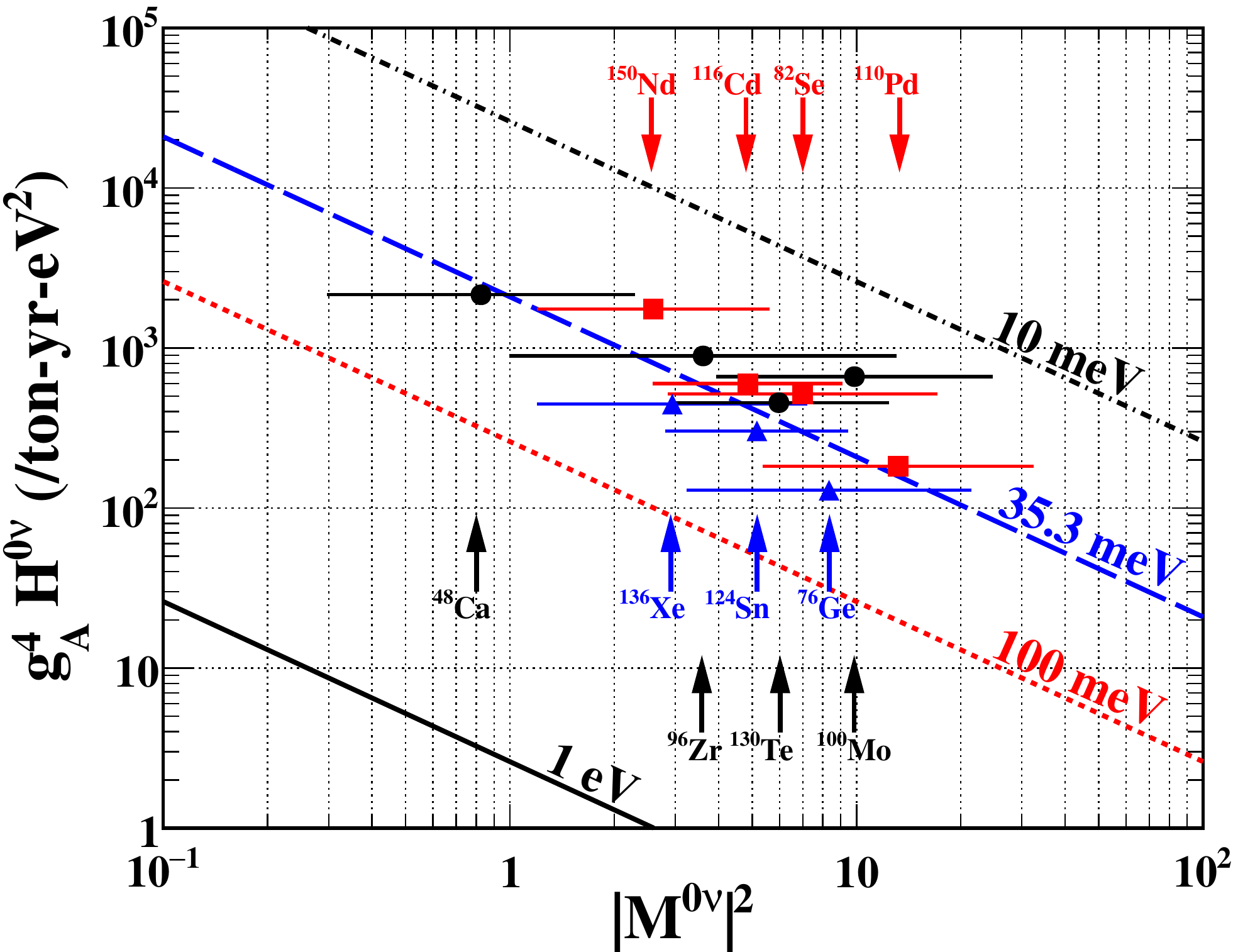}
\caption{
Variations of ``specific phase space''
$g_A^4 H^{0 \nu}$, as defined in Eq.~\ref{eq::H0nu},
versus $\ME2$ for various $\Abb$.
The geometric mean of the range of $\ME2$
is presented as the data points.
The best-fit and other diagonal lines correspond
to the matching $\mbb$ values at $g_A$=1.27
that give rise to a $\0nubb$ rate of 1 event per ton-year
at full efficiency.
This formulation is adopted from
Figures~2\&3 of Ref.~\cite{nubb-Robertson}.
}
\label{fig::modelfit}
\end{figure}


\begin{table}
\centering
\caption{
Summary of the key parameters used in this work.
Inputs are the IH and NH bands at $\pm 3 \sigma$ 
of the ND scenario at $m_{min} {<} 10^{-4} ~ {\rm eV}$
from existing measurements~\cite{nu-review},
such that \mbox{$\mbbmin {<} \mbb {<} \mbbmax$}. 
The posterior $\p95mbb$ denotes the 95\% lower bound 
for the $\mbb$-distribution, 
taking an uncorrelated $( \alpha , \beta )$ 
and the uncertainty range in $\ME2$
as prior~\cite{mbbprob}.
The corresponding minimal-exposures at 
background-free levels
under criteria $\P503s$
are given as $\Sigma_{\rm min}$.
The reduction fraction in $\Sigma$
from $\mbbmin$ to $\p95mbb$ is
denoted by $f_{95\%}$.
The values of $( \mbbmin , \mbbmax )$ and
$\p95mbb$ define the IH/NH band width  and dotted lines,
respectively, in 
Figures~\ref{fig::SigmaVsB0},\ref{fig::mbbVsBI}\&\ref{fig::SigmaVsmbb}
in this article.
}
\begin{center}
\renewcommand{\arraystretch}{1.1}
\begin{tabular}{|rcccc|}
\hline
& ~ $\mbbmin$  &  ~ $\mbbmax$  &  ~ $\p95mbb$  
& ~ $f_{95\%}$ ~ \\ \hline \hline 
\multicolumn{5}{|l|}{\bf IH:} \\
\multicolumn{5}{|l|}{~$\mbb ( {\times} 10^{\mbox{-}3} ~ {\rm eV} )$} \\
& 14 & 51 & 20 & $-$   \\
~ $\Sigma_{\rm min}$ (ton-yr) & 1.7 & 0.13 & 0.83   & 0.49  \\ \hline
\multicolumn{5}{|l|}{\bf NH:} \\
\multicolumn{5}{|l|}{~$\mbb ( {\times} 10^{\mbox{-}3} ~ {\rm eV} )$} \\
 & 0.78 & 4.3 & 3.0 &  $-$   \\
~ $\Sigma_{\rm min}$ (ton-yr) & 550 & 18 & 37  & 0.068 \\ \hline
\end{tabular}
\end{center}
\label{tab::IHNH}
\end{table}




Results from
neutrino oscillation experiments~\cite{nu-review,nuosc-review} 
indicate that the $m_i$ 
of the three active $\nu_i$ 
have structures corresponding to 
either IH or NH.
The values of $\mbb$ are constrained
and depend on the absolute neutrino mass scale,
and are typically expressed in terms of 
the lowest mass eigenstates $m_{min}$.
The $\pm 3 \sigma$ ranges of $\mbb$ with the 
non-degenerate (ND) mass eigenstate scenarios at 
$m_{min} {<} 10^{-4} ~ {\rm eV}$,
denoted by $\mbbmin$ and $\mbbmax$, are constant and 
listed in Table~\ref{tab::IHNH}.


There are no experimental
constraints on the Majorana phases $( \alpha , \beta )$.
It is in principle possible to have accidental cancellation
which leads to very small $\mbb$  at 
$m_{min} {\in} [1,10] {\times} 10^{-3} ~ {\rm eV}$.
However, under the reasonable assumption 
that they are uncorrelated and 
have uniform probabilities
within $[ 0 , 2 \pi ]$, a posterior probability distributions
of $\mbb$ can be assigned~\cite{mbbprob}.
The 95\% lower limit, denoted as $\p95mbb$ and listed in Table~\ref{tab::IHNH},
shows that the vanishing values of $\mbb$ are disfavored.



The current generation of oscillation experiments 
may reveal Nature's choice
between the two hierarchy options. 
In particular, there is an emerging preference of
NH over IH~\cite{nu-review,NHoverIH}.
Moreover, the combined cosmology data may provide 
a measurement on the sum of $m_i$~\cite{nucosmo-review}. 
Therefore, it can be expected that 
the ranges of parameter space of $\mbb$
in $\0nubb$ searches
will be further constrained.

Extracting neutrino mass information via Eq.~\ref{eq::thalf0nu} 
from the experimentally measured $\thalf0nu$
requires knowledge of $\ME2$ and $g_A$. 
There are different schemes to
calculate $\ME2$ for different $\Nbb$~\cite{dbdme-review}.
Deviations among their results are the main
contributors to the theoretical uncertainties.
Another source of uncertainties
is the values of $g_A$, which may 
differ between a free nucleon and complex nuclei~\cite{nubb-gA}.


Studies of Ref.~\cite{nubb-Robertson} suggest that,
in the case where $\0nubb$ is driven by the neutrino mass mechanism,
there exists an inverse correlation 
between $G^{0 \nu}$ and $\ME2$ in Eq.~\ref{eq::thalf0nu},
the consequence of which is that 
the decay rates per unit mass for different $\Nbb$
are similar at given $\mbb$ and constant $g_A$. 
That is, there is no favored $\0nubb$-isotope from the nuclear
physics point of view.

This empirical observation originates partially to the
large uncertainties in $\ME2$ and $g_A$.
To derive numerical results which would shed qualitative insights
without involving excessive discussions on the choice of $\ME2$,
we {\it assume} that this correlation is quantitatively valid.

We follow Ref.~\cite{nubb-Robertson} in adopting 
the geometric means of the realistic ranges
for the various $\ME2$ in different isotopes.
The data points can be parametrized by straight lines
at given $\mbb$, as depicted
in Figure~\ref{fig::modelfit}. 
That is, 
$[ \ME2 ( g_A^4 H^{0 \nu} ) ]$ is a constant at 
fixed $\mbb$ independent of $\Abb$.
The displayed $\mbb$ values in Figure~\ref{fig::modelfit}
correspond to $\0nubb$ decay rates of  
$[ N_{obs}^{0 \nu}/\Sigma ] {=} 1/\tyr$  
at $g_A$=1.27 and full efficiency.
The best-fit at this decay rate  corresponds to
$\mbb {=} ( 35 {\times} 10^{-3} ) ~ {\rm eV}$.


Following Eq.~\ref{eq::H0nu}, 
this model leads to a simplifying consequence that
\begin{equation}
\Sigma ( \tyr )~  \cdot ~ 
\left[ \frac{\varepsilon_{RoI}}{N_{obs}^{0 \nu}} \right]
~~ \propto ~~ \frac{ 1 }{ \mbb ^2 }   ~~  
\label{eq::nuphysconnect}
\end{equation}
at IA=100\% and $\varepsilon_{expt}$=100\%,
which is universally applicable to all $\Abb$.
The proportional constant can be derived via 
the best-fit values of Figure~\ref{fig::modelfit}.

Given a background $\B0$ as input,
the required $\S0$ to establish signal under $\P503s$
can be derived via Figure~\ref{fig::S0vsB0}a.
This is related to the mean of $N_{obs}^{0 \nu}$
at known $\varepsilon_{RoI}$.
Neutrino physics provides constraints on $\mbb$ with
several scales-of-interest  given in Table~\ref{tab::IHNH}.
The output values of $\Sigma$ and $\BI0$ can be derived
with Eqs.~\ref{eq::nuphysconnect}\&\ref{eq::BI}, respectively.

The $\Sigma$-values thus inferred 
in what follows could be interpreted with
the typical uncertainties of a ``factor of two, both directions''
(that is, within a factor of $[ 0.5, 2.0 ]$ of the nominal values)
to match our current understanding
of $\ME2$.



\section{Sensitivity Dependence}

It is well-known, following
Eqs.~\ref{eq::thalf0nu}\&\ref{eq::roi},
that
the sensitivity to $\left[ 1/\mbb \right] $ is  proportional to
$\Sigma ^{\scaleto{\frac{1}{2}}{8pt}}$ as $\B0 {\rightarrow} 0$
and to $\Sigma ^{\scaleto{\frac{1}{4}}{8pt}}$ at large $\B0$.
We further investigate the $\B0$-dependence quantitatively
and in the context of the preferred IH and NH ranges
with the model of
Ref.~\cite{nubb-Robertson}.
The specific $\mbb$-values of Table~\ref{tab::IHNH}
$-$ ($\mbbmin , \mbbmax , \p95mbb$) for both IH and NH $-$
serve to provide reference scales.



\begin{figure}
\includegraphics[width=8.2cm]{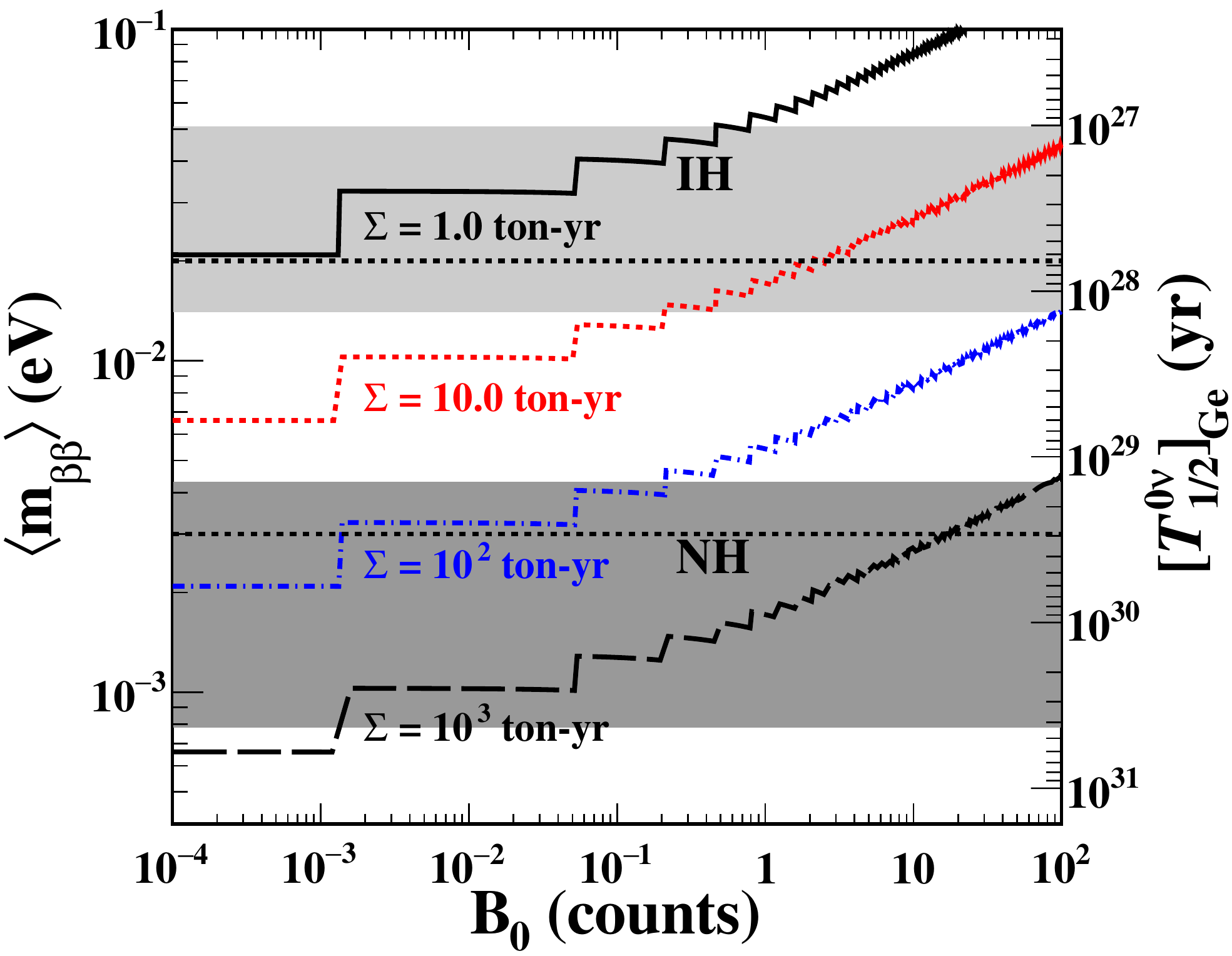}
\caption{
The variation of $\mbb$ with $\B0$ following $\P503s$,
with RoI=$\FWHM$ for $\ge76$ at 
$\Sigma {=} ( 1 ; 10 ; 100 ; 1000 ) ~ \tyr$
following Eq.~\ref{eq::H0nu} at $g_A$=1.27.
The IH(NH) bands are defined by
$( \mbbmin , \mbbmax )$, while 
their $\p95mbb$-values are denoted as dotted lines.
The corresponding variations of
$\left[ \thalf0nu \right] _{\rm Ge}$ versus $\B0$ 
 are displayed as the right vertical axis.
This is specific to $\ge76$, while the equivalent
values for other $\Nbb$ can be derived via Eq.~\ref{eq::Ge-Convert}.
}
\label{fig::SigmaVsB0}
\end{figure}


\begin{figure}
\includegraphics[width=8.2cm]{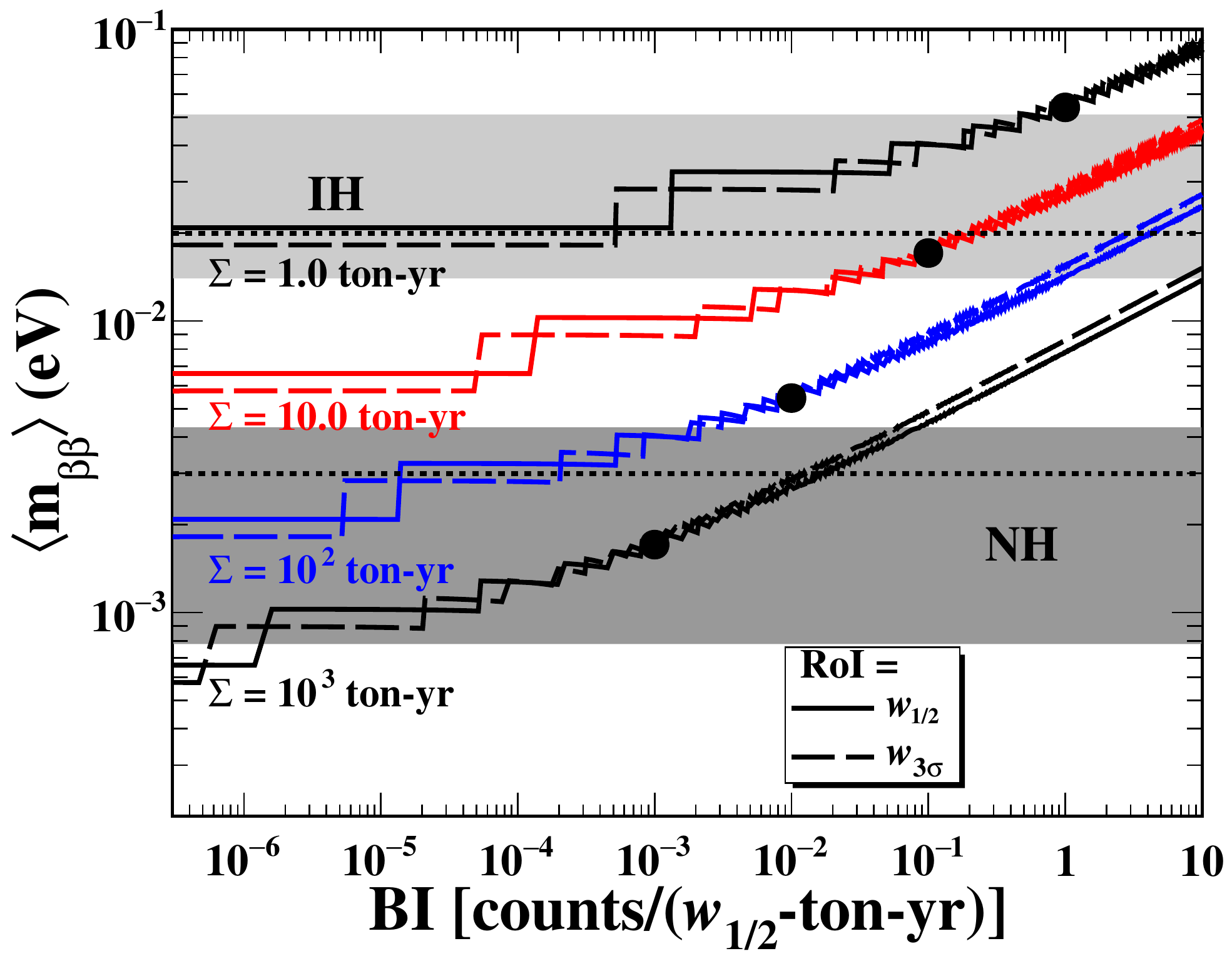}
\caption{
Sensitivities of $\mbb$ 
versus $\BI0$, as defined in Eq.~\ref{eq::BI}
which is universally applicable to all $\Abb$,
following $\P503s$
under different exposures
at $\Sigma {=} ( 1 ; 10 ; 100 ; 1000 ) ~ \tyr$.
The IH(NH) bands are defined by
$( \mbbmin , \mbbmax )$, while 
their $\p95mbb$-values are denoted as dotted lines.
Both options of RoI=$\FWHM$ and $\FWFM$ are displayed.
Black dots correspond to the 
benchmark [1~$\BM$] background levels.
}
\label{fig::mbbVsBI}
\end{figure}


\begin{figure}
{\bf (a)}\\
\includegraphics[width=8.2cm]{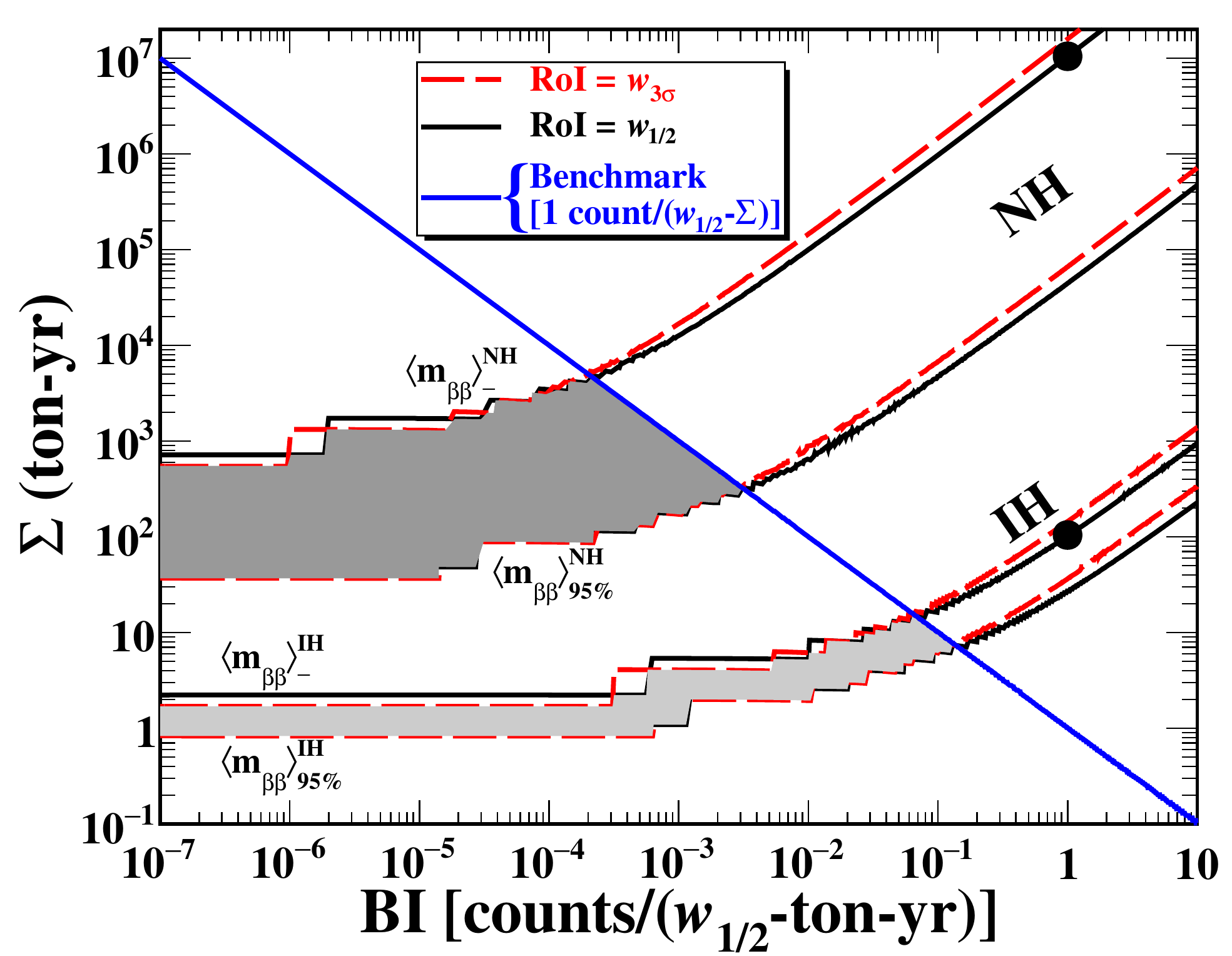}\\
{\bf (b)}\\
\includegraphics[width=8.2cm]{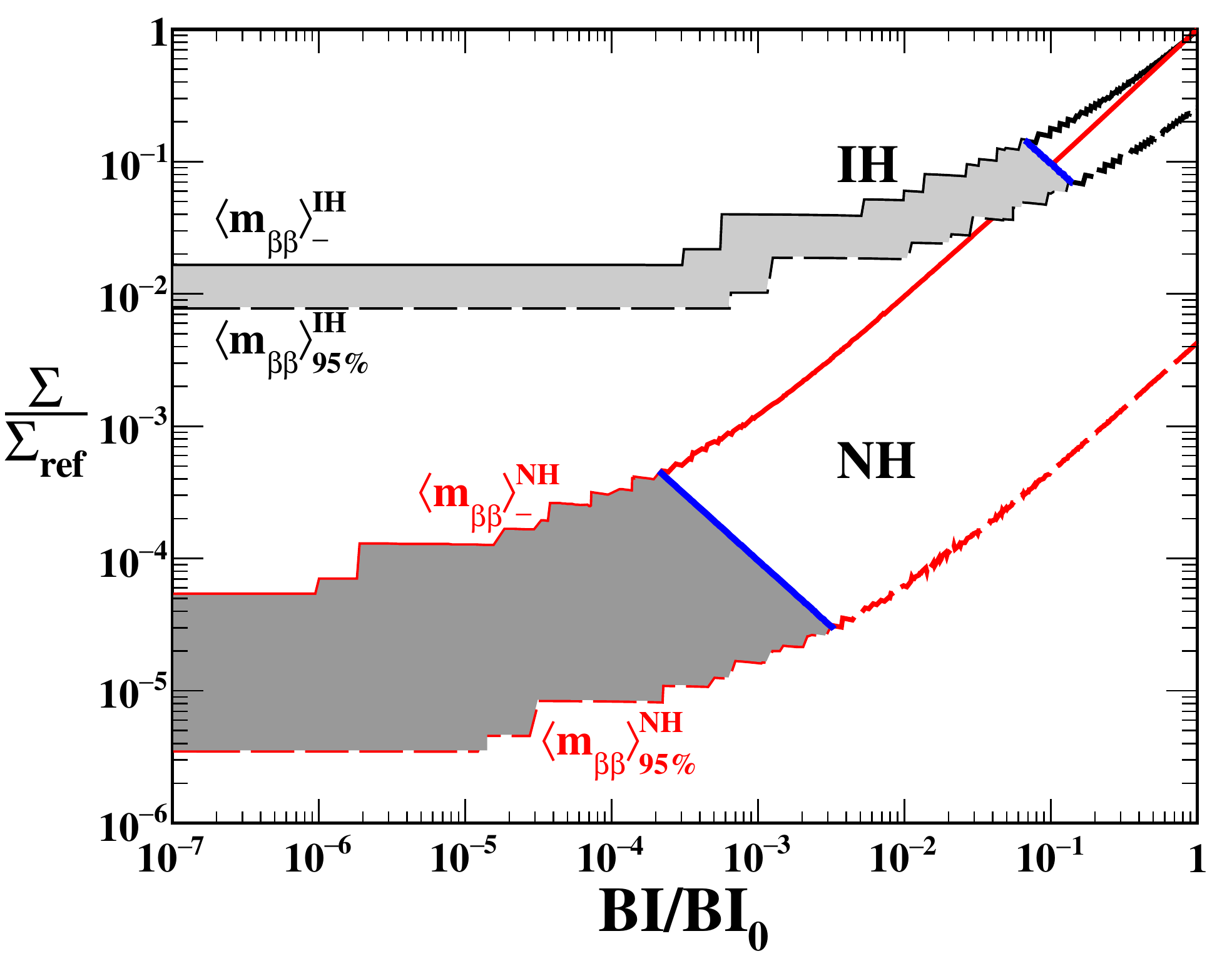}
\caption{
(a) Required exposure
universally applicable to all $\Nbb$
to cover $\p95mbb$ and $\mbbmin$ under $\P503s$
at IH and NH
versus $\BI0$. 
The benchmark background [1~$\BM$] condition is
superimposed as the blue contour.
The shaded regions correspond to the preferred
hardware specification space for future $\0nubb$ experiments.
(b) The relation between
exposure and background reduction
relative to the current $\BIcurrent {=} 1 {~} \Bnowunit$  
and its corresponding required exposures
$\Sigma_{\rm ref}^{\rm IH(NH)} {=} {\rm 110~\tyr
( 11~M\tyr )}$ for IH(NH).
The reference points $ ( \BIcurrent , \Sigma_{\rm ref}^{\rm IH(NH)} )$ are
represented by black dots in (a). 
Sensitivities with both RoI=$\FWHM$ and $\FWFM$ are displayed in (a),
while the more sensitive of the two schemes are shown in (b).
The shaded regions match those of (a).
}
\label{fig::SigmaVsBI}
\end{figure}


\begin{figure}
\includegraphics[width=8.2cm]{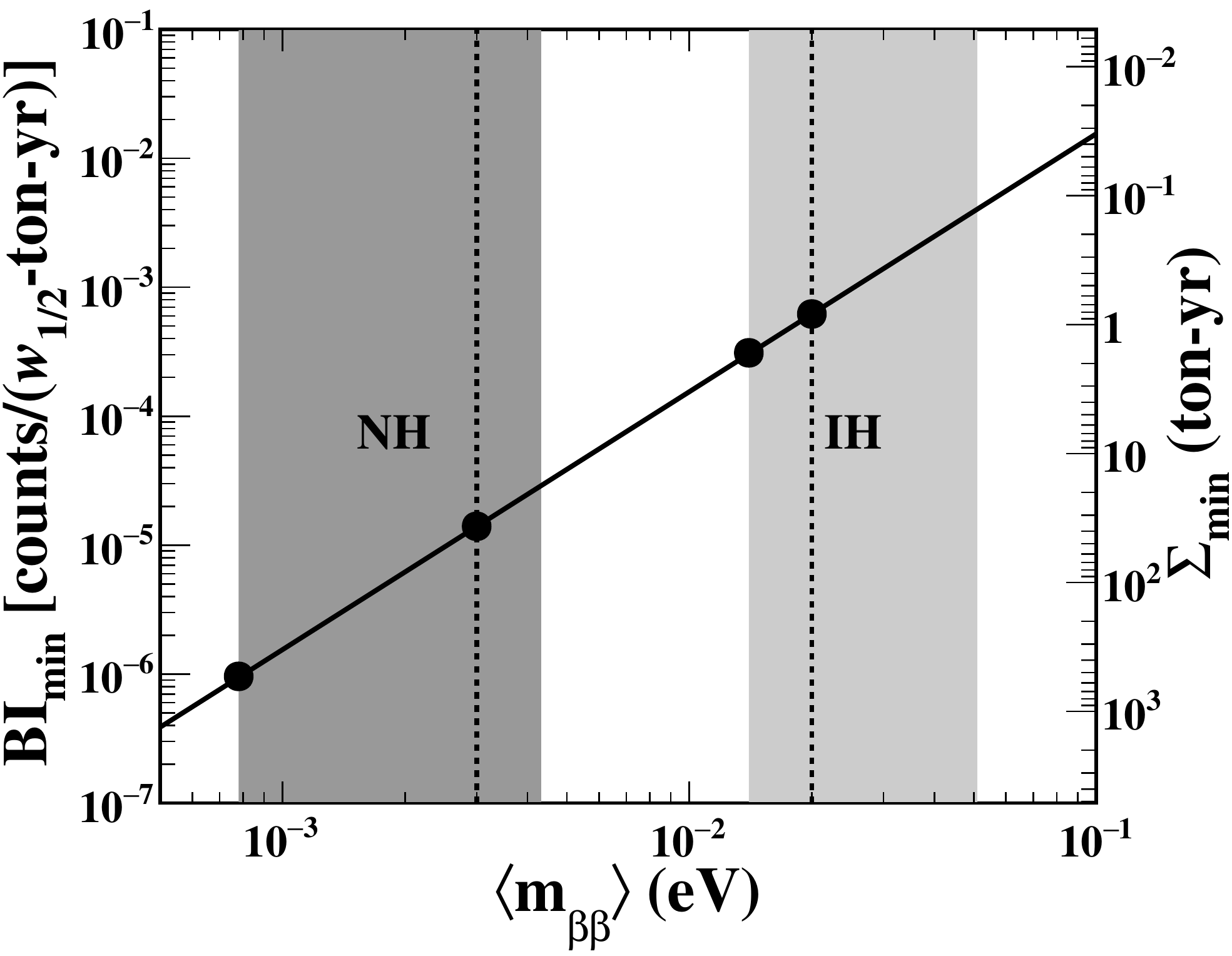}\\
\caption{
The variations of the
background-free and minimal-exposure conditions
$( \BI0 _{\rm min} , \Sigma_{\rm min} )$
at $\P503s$-criteria with $\mbb$.
The IH(NH) bands are defined by
$( \mbbmin , \mbbmax )$, while 
their $\p95mbb$-values are denoted as dotted lines.
The black dots correspond to the values listed
in the last rows of Table~\ref{tab::bkgscenario}.
}
\label{fig::bkgfree}
\end{figure}


\begin{figure}
{\bf (a)}\\
\includegraphics[width=8.2cm]{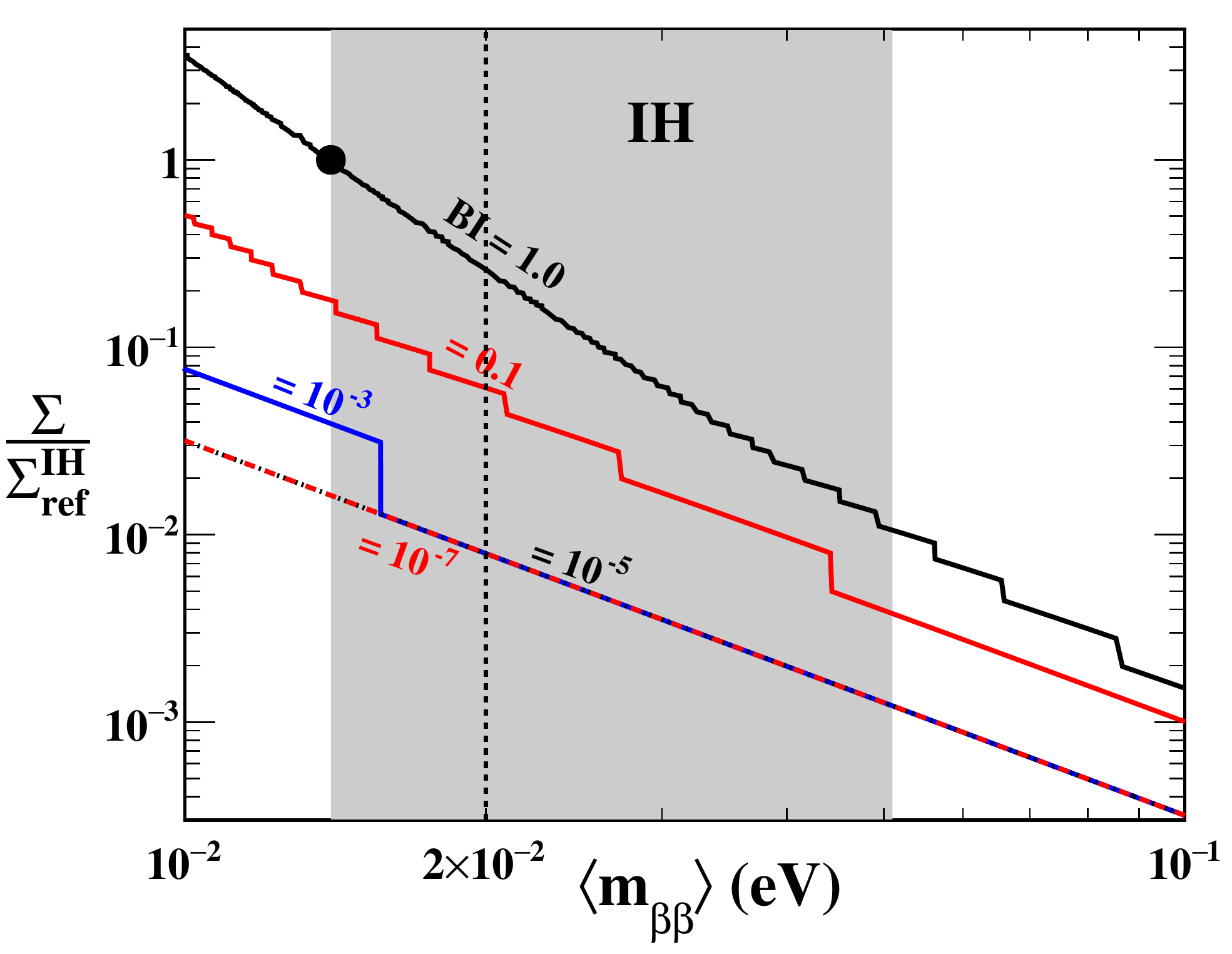}\\
{\bf (b)}\\
\includegraphics[width=8.2cm]{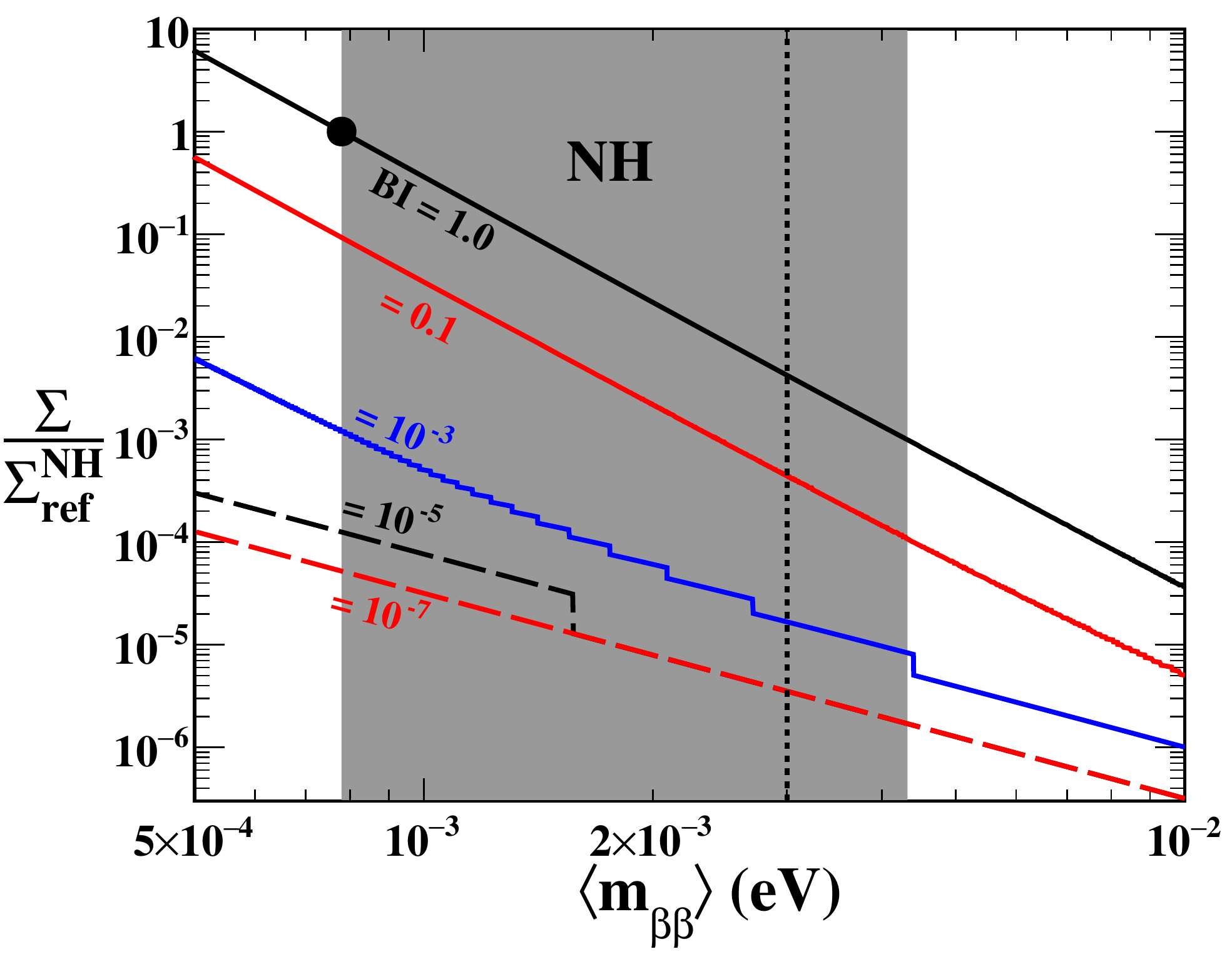}
\caption{
Fractional reduction of required exposure
to probe (a) IH and (b) NH
as a function of $\mbb$ under
different $\BI0$ background levels
in unit of [$\BIunit$].
The IH(NH) bands are defined by
$( \mbbmin , \mbbmax )$, while 
their $\p95mbb$-values are denoted as dotted lines.
The reference exposures
$\Sigma_{\rm ref}^{\rm IH(NH)} {=} {\rm 110~\tyr ( 11~M\tyr )}$ 
to cover $\mbbmin ^{\rm IH(NH)}$, denoted as black dots,
correspond to those required under the
current background level of $\BIcurrent {=} 1 {~} \Bnowunit$.
}
\label{fig::SigmaVsmbb}
\end{figure}


\subsection{Required Exposure and Background}
\label{sect::duality}


\begin{table*}
\centering
\caption{
Required exposure ($\Sigma$) 
to cover $\p95mbb^{\rm IH(NH)}$ and $\mbbmin^{\rm IH(NH)}$ at
different background scenarios in descending order of intensity.
The BI-values follow from Eq.~\ref{eq::BI}.
}
\begin{center}
\renewcommand{\arraystretch}{1.1}
\begin{tabular}{|lc||cc|cc|}
\hline
\multicolumn{2}{|c||}{Background}   & 
\multicolumn{4}{c|}{Required $\Sigma$ (ton-yr) To Cover} \\ \cline{3-6} 
\multirow{2}{*}{Scenario} & 
\multirow{2}{*}{$\BI0$} & 
\multirow{2}{*}{$~ \p95mbb^{IH} ~$} & \multirow{2}{*}{$~ \mbbmin^{IH} ~$} &
\multirow{2}{*}{$~ \p95mbb^{NH} ~$} & \multirow{2}{*}{$~ \mbbmin^{NH} ~$} \\
& [$\BIunit$] & & & & \\ \hline \hline
Best Published~\cite{GERDA} & 1 &
27 & 110 & $4.4 {\times} 10^4 $  & $ 11 {\times} 10^6 $  \\
Next Generation  \multirow{2}{*}{ $ \} $ } & 
\multirow{2}{*}{ 0.1} & 
\multirow{2}{*}{ 6.1 } & 
\multirow{2}{*}{19} & 
\multirow{2}{*}{$4.7 {\times} 10^3$ } & 
\multirow{2}{*}{$0.97 {\times} 10^6$}  \\
~~Projected~\cite{NG0nubb} & & & & &   \\ 
Benchmark  
\multirow{2}{*}{ $ {\rm ~~~~~~~~~~~~~~ IH:} \{ $ } & 
0.14 & 7.3 & -- & -- & -- \\
~~ [$1~\BM$]  & 0.067 & -- & 15 & -- & -- \\
\multirow{2}{*}{$ {\rm \hspace*{2cm} ~~~~~~~~~~~ NH:} \{ $ } &
$3.1 {\times} 10^{-3}$ & -- & -- & 330 & -- \\  
& $2.2 \times 10^{-4}$ & -- & --  & -- & $4.6 {\times} 10^3$  \\
``Background \multirow{2}{*}{ $ {\rm ~~~~~~~~~~~ IH:} \{ $ } & 
${\leq} ( 6.3 {\times} 10^{-4} ) $ & 
0.83 & --  & -- & -- \\
~~ -Free''  &  
${\leq} ( 3.1 {\times} 10^{-4} ) $ & 
0.83 & 1.7 & -- & -- \\
\multirow{2}{*}{$ {\rm \hspace*{2cm} ~~~~~~~~~~~ NH:} \{ $ } & 
${\leq} ( 1.4 {\times} 10^{-5} )$ &
0.83 & 1.7 & 37 & -- \\   
& ${\leq} ( 0.96 \times 10^{-6} )$ & 
0.83 & 1.7  & 37 & 550  \\ \hline
\end{tabular}
\end{center}
\label{tab::bkgscenario}
\end{table*}



\begin{table*}
\caption{
The range of $( \S0 , \B0 )$ to qualify a positive signal
to cover $\mbbmin$ for both IH and NH under $\P503s$, 
given the observed number of
events in RoI $-$ $\0nubb$-signals and background are 
combined but indistinguishable at event-by-event level.
The smaller $\Sigma$ values 
among the alternatives of 
RoI=$\FWHM$ or $\FWFM$ 
are selected.
The sixth column shows the required 
$\BI0$ 
which is universal to all $\Nbb$.
Last column lists the required background
specifically for  $\ge76$
normalized to  ``$\kty$'', and the conversion to
other isotopes is referred
to Eq.~\ref{eq::conversion}.
The $N_{obs}^{0 \nu} {=} 1$ row correpsonds to the
background-free conditions.
The BI-values follow from Eq.~\ref{eq::BI}.
}
\label{tab::SigmaVsBI}
\centering
\begin{center}
\begin{tabular}{|ccccccc|}
\hline

\multicolumn{3}{|c}{ Counts Within RoI }
& Optimal           & Required        & Universal     & Background$\kty$ \\
\multirow{2}{*}{$N_{obs}^{0 \nu}$}    
& \multirow{2}{*}{$\S0$}   &  \multirow{2}{*}{$\B0$}
& RoI                  & Exposure          & $\BI0$   &  for $^{76}$Ge at \\
& &  & & $\Sigma$ (ton-yr)     &              
 $[ \BIunit ]$       &   $\Delta$ = 0.12\%   \\ \hline \hline

\multicolumn{7}{|l|}{ }  \\
\multicolumn{7}{|l|}{ \underline{Covering $\mbbmin$ for IH:} }  \\

    1        & $\geq$ 0.69   & ~ $\leq$ 1.3$\times10^{-3}$ ~    &~ $\FWFM$ ~                  & 1.7              & ~ $\leq$ 3.1$\times10^{-4}$ ~   & $\leq$ 1.2$\times10^{-4}$     \\ 
    2        & $\geq$ 1.6    & $\leq$ 5.2$\times10^{-2}$    &$\FWFM$                  & 4.0              & $\leq$ 5.2$\times10^{-3}$   & $\leq$ 2.1$\times10^{-3}$     \\ 
    3        & $\geq$ 2.5    & $\leq$ 0.21                 &$\FWFM$                  & 6.0              & $\leq$ 1.4$\times10^{-2}$   & $\leq$ 5.6$\times10^{-3}$      \\
    4        & $\geq$ 3.2    & $\leq$ 0.45                 &$\FWHM$                  & 10               & $\leq$ 4.3$\times10^{-2}$   & $\leq$ 1.8$\times10^{-2}$      \\
    5        & $\geq$ 3.9    & $\leq$ 0.77                 &$\FWHM$                  & 13               & $\leq$ 6.1$\times10^{-2}$   & $\leq$ 2.4$\times10^{-2}$      \\
   10       & $\geq$ 6.6    & $\leq$ 3.1                  &$\FWHM$                  & 21               & $\leq$ 0.14                & $\leq$ 6.0$\times10^{-2}$      \\ \hline
   
\multicolumn{7}{|l|}{ }  \\
\multicolumn{7}{|l|}{ \underline{Covering $\mbbmin$ for NH:} }  \\

   1         & $\geq$ 0.69   & $\leq$ 1.3$\times10^{-3}$    &$\FWFM$                  & 5.5$\times10^{2}$ & $\leq$ 0.96$\times10^{-6}$  & $\leq$ 0.38$\times10^{-6}$     \\       
   2         & $\geq$ 1.6    & $\leq$ 5.2$\times10^{-2}$    &$\FWFM$                  & 1.3$\times10^{3}$ & $\leq$ 1.6$\times10^{-5}$   & $\leq$ 6.4$\times10^{-6}$       \\       
   3         & $\geq$ 2.5    & $\leq$ 0.21                 &$\FWFM$                  & 2.0$\times10^{3}$ & $\leq$ 4.0$\times10^{-5}$   & $\leq$ 1.7$\times10^{-5}$       \\       
   4         & $\geq$ 3.2    & $\leq$ 0.45                 &$\FWHM$                  & 3.4$\times10^{3}$ & $\leq$ 1.3$\times10^{-4}$   & $\leq$ 5.4$\times10^{-5}$       \\       
   5         & $\geq$ 3.9    & $\leq$ 0.77                 &$\FWHM$                  & 4.2$\times10^{3}$ & $\leq$ 1.8$\times10^{-4}$   & $\leq$ 7.5$\times10^{-5}$       \\       
   10        & $\geq$ 6.6    & $\leq$ 3.1                  &$\FWHM$                  & 7.0$\times10^{3}$ & $\leq$ 4.4$\times10^{-4}$   & $\leq$ 1.8$\times10^{-4}$       \\ \hline

\end{tabular}
\end{center}
\end{table*}



The variations of $\mbb$ versus $\B0$ 
with different $\Sigma$ at 
RoI=$\FWHM$ (such that $\varepsilon_{RoI} {\simeq} 76\%$)
under the criteria of $\P503s$
are depicted in Figure~\ref{fig::SigmaVsB0}, 
with the IH and NH bands superimposed.
The matching $ \thalf0nu $ for $\ge76$ is illustrated. 
The equivalent half-life sensitivities for 
other isotopes $\Nbb$
can be derived via 
\begin{equation}
\left[ \thalf0nu \right] _{\Abb} = 
\left[ \thalf0nu \right] _{\ge76}  \left( \frac{76}{\Abb} \right)  ~.
\label{eq::Ge-Convert}
\end{equation}
The figure depicts how the same exposure 
can be used to probe longer $ \thalf0nu $ 
and smaller $\mbb$ with decreasing background.


The dependence of $\mbb$ sensitivities to $\BI0$ 
is depicted in Figure~\ref{fig::mbbVsBI}.
Taking RoI=$\FWHM$  is obviously not the optimal choice
when the expected background $\B0 {\rightarrow} 0$.
An alternative choice for low $\B0$ 
is RoI${\equiv} \FWFM$ covering $\pm 3 \sigma$ of $\Qbb$,
such that $\varepsilon_{RoI} {\cong} 100\%$.
Both schemes are illustrated in 
Figure~\ref{fig::mbbVsBI}.
The choice of
RoI=$\FWFM$ at $\B0 {\rightarrow} 0$
would expectedly give better sensitivity
by a factor of $\varepsilon_{RoI}$($\FWHM$)=0.76,
such that the covered $\thalf0nu$ is 32\% longer,
or the required $\Sigma$ is 24\% less.


The required exposure to probe $\p95mbb$ and
$\mbbmin$ with both RoI selections 
in both IH and NH are depicted 
in Figure~\ref{fig::SigmaVsBI}a.
Superimposed as a blue contour is the
``benchmark'' background level at 1~$\BM$ 
where the first background event 
would occur at a given exposure. 
The benchmark level also represents 
the transition in the effectiveness of probing 
$\mbb$ with increasing exposure.
The shaded regions correspond to the
preferred hardware specification space
for future $\0nubb$ experiments $-$ where
the exposure should be sufficient to cover
at least $\p95mbb ^{\rm IH(NH)}$,  
and there would be less than one background event
per $\FWHM$ over the full exposure.


The required exposures 
under various background conditions 
are summarized in Table~\ref{tab::bkgscenario}.
The best published background level
is $1.0^{+0.6}_{-0.4}~{\rm counts}\kty$  
or $\BI0 {\sim} 3 {~} \BIunit$ 
from the GERDA experiment on $\ge76$~\cite{GERDA}.
For simplicity, 
the ``best'' current background is taken to be 
$\BI0 {\equiv} \BIcurrent {=} 1 {~} \Bnowunit$ 
in what follows.
This background would correspond to
$\Sigma_{\rm ref}^{\rm IH(NH)} {=} {\rm 110~\tyr ( 11~M\tyr )}$
to cover $\mbbmin^{\rm IH} ( \mbbmin^{\rm NH} )$.

Such a large required exposure is inefficient and unrealistic, 
so that the background should be significantly reduced 
to allow the quest to advance.
The target exposure is $\Sigma$=10~ton-yr
for the next generation $\0nubb$ projects  to cover IH 
with ton-scale detector target~\cite{NG0nubb}.
Following Figure~\ref{fig::mbbVsBI},
this exposure would require
$\BI0 {<} ( 0.21 , 0.033 ) {~} \BIunit$ to
cover $( \p95mbb , \mbbmin^{\rm IH} )$.
This  matches the background specifications of
$\BI0 {=} \mathcal{O}(0.1) {~} \BIunit$.   


The background-free ($\BI0 _{\rm min}$)
$-$ equivalently, minimal-exposure ($\Sigma _{\rm min}$) $-$ condition
is where one single observed event can establish
the signal at the $\P503s$-criteria.
Their values at the benchmark $\mbb$'s are given
in Table~\ref{tab::IHNH}. 

The choice of $\mbbmin$ to define $\Sigma$
is a conservative one.
Since \mbox{$\p95mbb {>} \mbbmin$} from Table~\ref{tab::IHNH},
the minimum exposure $\Sigma_{\rm min}$
corresponding to $\p95mbb$ is reduced relative to
that for $\mbbmin$ by a fraction 
given as $f_{95\%}$.

The variations of $( \BI0 _{\rm min} , \Sigma_{\rm min} )$
with $\mbb$ are depicted in Figure~\ref{fig::bkgfree}.
As shown by the black dots and also listed in Table~\ref{tab::bkgscenario}, 
$\Sigma_{\rm min}$=(0.83,1.7)~ton-yr at 
$\BI0 _{\rm min} {\leq} ( 6.3 {\times} 10^{-4} , 3.1 {\times} 10^{-4} ) {~} \BIunit$
are required to cover $( \p95mbb , \mbbmin )^{\rm IH}$.
The corresponding requirements for NH are 
$\Sigma_{\rm min}$=(37,550)~ton-yr at
$\BI0 _{\rm min} {\leq} ( 1.4 {\times} 10^{-5} , 0.96 {\times} 10^{-6} ) {~} \BIunit$.
The required $\Sigma_{\rm min}$ from $\mbbmin$ to $\p95mbb$ 
is reduced by  $f_{95\%}$=0.49(0.068) for IH(NH).

Alternatively, the $\Sigma$=10~ton-yr target exposure of next-generation projects 
can probe $\mbb {>} ( 5.8 {\times} 10^{-3} )~{\rm eV}$, 
approaching $\mbbmax^{\rm NH} {=} ( 4.3 {\times} 10^{-3} )~{\rm eV}$, 
when the background-free condition
$\BI0 _{\rm min} {<} 5.1 {\times} 10^{-5} {~} \BIunit$
is achieved.

The interplay between fractional reduction of $\BI0$ and $\Sigma$
relative to $\BIcurrent$ and $\Sigma_{\rm ref}^{\rm IH(NH)}$ 
to cover $\p95mbb$ and $\mbbmin$ in IH(NH)
is depicted in Figure~\ref{fig::SigmaVsBI}b.
Background-free conditions require additional 
$\BI0$-suppression by factors of  
$3.1 \times 10^{-4}$($0.96 \times 10^{-6}$),
to cover $\mbbmin^{\rm IH(NH)}$
in which cases $\Sigma$ can be reduced
by factors of 0.016($5 \times 10^{-5}$). 
The shaded regions match those of Figure~\ref{fig::SigmaVsBI}a
in displaying the preferred hardware specification space.


The impact of background suppression to the required exposure is 
increasingly enhanced as smaller values of $\mbb$ are probed.
This is illustrated in Figures~\ref{fig::SigmaVsmbb}a(b)
which display the reduction fraction in $\Sigma$ relative
to $\Sigma_{\rm ref}^{\rm IH(NH)}$ at different background levels.
For instance, the suppression of $\BI0$ from 1 to $10^{-3} {~} \BIunit$
will contribute to the reduction of $\Sigma$ from
$\Sigma {=} (27,110) ~ \tyr$ to (1.1,4.1)~ton-yr
and from
$\Sigma {=} (44,11000) ~ {\rm k\tyr}$ to (0.17,13)~kton-yr 
to cover $( \p95mbb , \mbbmin )^{\rm IH}$ 
and $( \p95mbb , \mbbmin )^{\rm NH}$, respectively.


In realistic experiments, 
signals and background are indistinguishable
at the event-by-event level.
The expected average background $\B0$ and 
the observed event counts (an integer)
in the RoI are the known quantities.
They can be used to assess 
whether a signal is ``established'' under certain
criteria like $\P503s$.
Listed in Table~\ref{tab::SigmaVsBI}
are the required 
ranges on $( \S0 , \B0 )$
to qualify positive signals
given the number of observed events.
The first row corresponds to the background-free
condition, in which
one single event is sufficient to establish a signal.
Accordingly, the $( \BI0 , \Sigma )$ values 
match the entries in the last rows 
of Table~\ref{tab::bkgscenario},
and are displayed in Figure~\ref{fig::bkgfree}.

Results of Table~\ref{tab::SigmaVsBI},
apply generically to all $\Nbb$ except
those for the last column when background
is expressed in ``$\kty$'' unit.
The values are specific for $\ge76$, where 
the best published $\Delta ( \ge76 )$=0.12\%
of the MJD-experiment~\cite{MJD} is adopted as input.
The background requirements
for other $\Nbb$ can be derived via:
\begin{eqnarray}
\frac{[ {\rm Background}\kty  ] ( \Nbb)}
{[ {\rm Background}\kty ] ( \ge76 )} \hspace*{2cm} \nonumber \\
 ~~~~~ =
\left[ \frac{  \Delta (\ge76 ) }{ \Delta ( \Nbb ) } \right]
\left[ \frac{ \Qbb ( \ge76 ) }{ \Qbb ( \Nbb ) } \right] .
\label{eq::conversion}
\end{eqnarray}


\begin{figure}
\includegraphics[width=8.2cm]{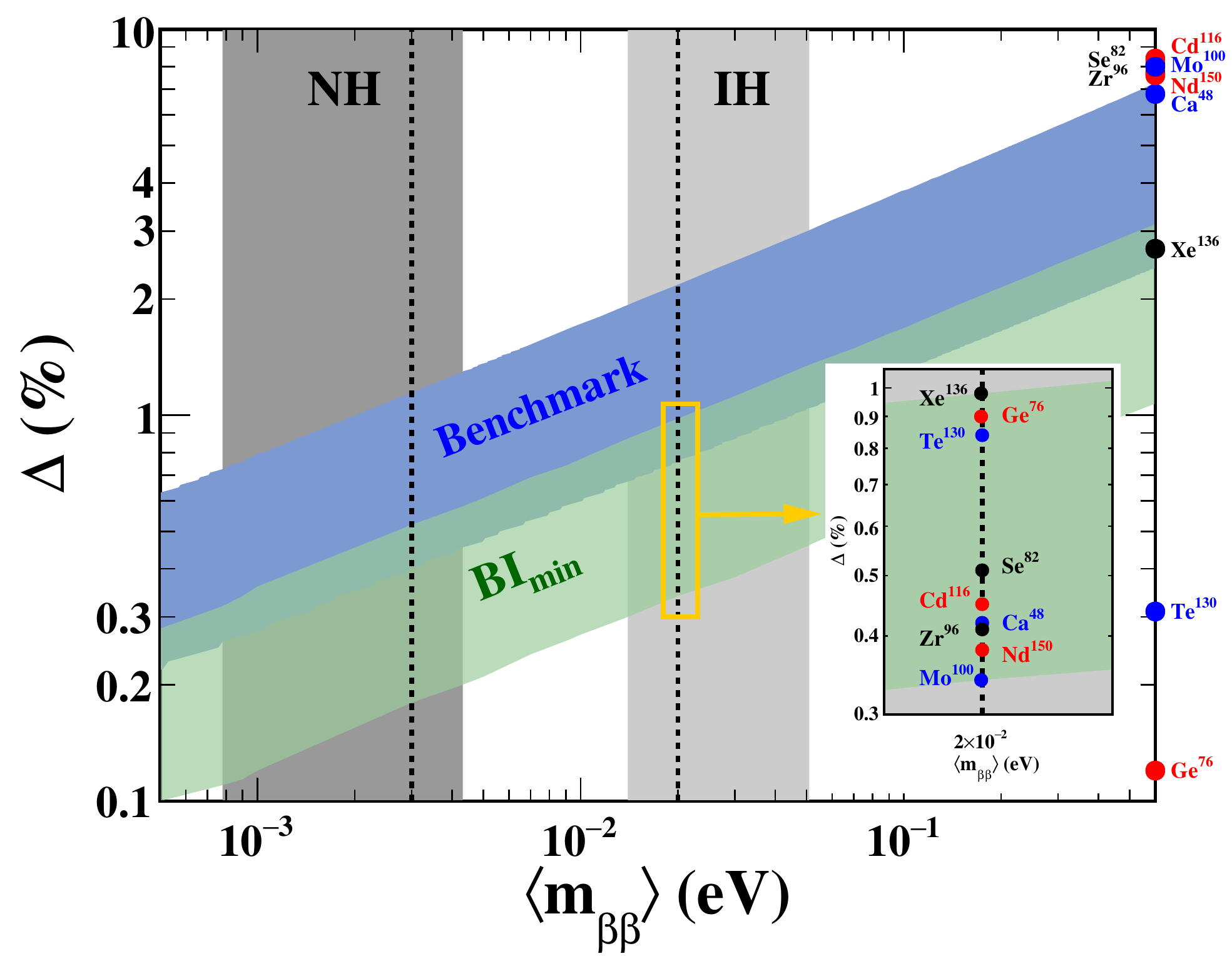}\\
\caption{
Variations of the required $\Delta$ with $\mbb$
such that $\2nubb$ background within RoI=$\FWFM$ would contribute
less than those specified by 
the benchmark [1~$\BM$] level and 
the background-free ($\BI0 _{\rm min}$)
condition of Figure~\ref{fig::bkgfree} as 
a function of $\mbb$.
The relative locations of $\Abb$ within both
bands are depicted in the inset, 
using the $\BI0 _{\rm min}$-band as illustration.
The best achieved $\Delta$'s in past and ongoing
experiments are displayed at the right vertical axis. 
}
\label{fig::Delta2nubb}
\end{figure}


 \begin{table*}
\centering
\caption{
The required $\Delta$ for selected $\Nbb$,
listed in descending order of 
their measured $\tauhalf^{2\nu}$~\cite{nubb-review,GERDA},
such that the $\2nubb$ background within RoI=$\FWFM$ 
would contribute less than the levels specified by
the benchmark and background-free conditions
to cover $\mbb_{95\%}^{\rm IH(NH)}$ and
$\mbbmin^{\rm IH(NH)}$. 
The BI-values follow from Eq.~\ref{eq::BI}.
}
\begin{center}
\renewcommand{\arraystretch}{1.1}
\begin{tabular}{|cccc||cccc|cccc|}
\hline
\multicolumn{4}{|c||}{}   & 
\multicolumn{4}{c|}{\underline{Inverted Hierarachy}} &  
\multicolumn{4}{c|}{\underline{Normal Hierarachy}}  \\
\multicolumn{4}{|c||}{}   & 
\multicolumn{4}{c|}{$\BI0 ~ [ \BIunit ] $} &  
\multicolumn{4}{c|}{$\BI0 ~ [ \BIunit ] $}  \\ 
\multicolumn{4}{|c||}{Sensitivity:}   & 
\multicolumn{2}{c}{Benchmark } &
\multicolumn{2}{c|}{Background-Free }   & 
\multicolumn{2}{c}{Benchmark }  &
\multicolumn{2}{c|}{Background-Free }   \\ \hline
\multicolumn{4}{|c||}{}   & & & & & & & & \\
\multicolumn{4}{|c||}{$\leq \p95mbb$} 
&  ${\leq} 0.14$  & $-$ 
& ${\leq} 6.3 {\times} 10^{-4}$ & $-$
& ${\leq} 3.1 {\times} 10^{-3}$  & $-$ 
& ${\leq} 1.4 {\times} 10^{-5}$ & $-$  \\
\multicolumn{4}{|c||}{$\leq \mbbmin$} 
& $-$ &  ${\leq} 6.7 {\times} 10^{-2}$ 
& $-$ &  ${\leq} 3.1 {\times} 10^{-4}$  
& $-$ &  ${\leq} 2.2 {\times} 10^{-4}$ 
& $-$ &  ${\leq} 0.96 {\times} 10^{-6}$  \\ \hline
$\Abb$ & $\Qbb$ & $\tauhalf^{2\nu}$ 
& Best$^\dagger$  &  \multicolumn{8}{c|}{}  \\ 
& (MeV)  & (yr)  & $\Delta (\%)$ &
\multicolumn{8}{c|}{Required $\Delta$ (\%) }  \\ 
\hline \hline

\multicolumn{4}{|c||}{}   & & & & & & & & \\


$^{136}$Xe  &  2.458         & $2.2 {\times} 10^{21}$   
& 2.7
& $\leq$ 2.18  & $\leq$ 1.93 & $\leq$ 0.98 & $\leq$ 0.87 
& $\leq$ 1.15  & $\leq$ 0.73 & $\leq$ 0.52 & $\leq$ 0.32  \\      

$^{76}$Ge   &  2.039         & $1.9 {\times} 10^{21}$   
& 0.12
& $\leq$ 1.99  & $\leq$ 1.76 & $\leq$ 0.90 & $\leq$ 0.79 
& $\leq$ 1.04  & $\leq$ 0.67 & $\leq$ 0.46 & $\leq$ 0.30  \\      

$^{130}$Te  &  2.528         & $8.2 {\times} 10^{20}$   
& 0.31
& $\leq$ 1.89  & $\leq$ 1.67 & $\leq$ 0.84 & $\leq$ 0.74 
& $\leq$ 0.99  & $\leq$ 0.63 & $\leq$ 0.44 & $\leq$ 0.28  \\     

$^{82}$Se   &  2.998         & $9.2 {\times} 10^{19}$   
& 8.1
& $\leq$ 1.15  & $\leq$ 1.02 & $\leq$ 0.51 & $\leq$ 0.46 
& $\leq$ 0.61  & $\leq$ 0.39 & $\leq$ 0.27 & $\leq$ 0.18  \\    

$^{48}$Ca   &  4.268         & $6.4 {\times} 10^{19}$   
& 6.8
& $\leq$ 0.93  & $\leq$ 0.82 & $\leq$ 0.42 & $\leq$ 0.37 
& $\leq$ 0.49  & $\leq$ 0.31 & $\leq$ 0.22 & $\leq$ 0.14  \\  

$^{116}$Cd  &  2.814         & $2.7 {\times} 10^{19}$   
& 8.4
& $\leq$ 0.99  & $\leq$ 0.88 & $\leq$ 0.45 & $\leq$ 0.40 
& $\leq$ 0.52  & $\leq$ 0.34 & $\leq$ 0.23 & $\leq$ 0.15  \\ 

$^{96}$Zr   &  3.350         & $2.4 {\times} 10^{19}$   
&  7.7
& $\leq$ 0.91  & $\leq$ 0.81 & $\leq$ 0.41 & $\leq$ 0.36 
& $\leq$ 0.48  & $\leq$ 0.31 & $\leq$ 0.22 & $\leq$ 0.14  \\

$^{150}$Nd  &  3.371         & $9.3 {\times} 10^{18}$   
& 7.6
& $\leq$ 0.84  & $\leq$ 0.75 & $\leq$ 0.38 & $\leq$ 0.33 
& $\leq$ 0.45  & $\leq$ 0.29 & $\leq$ 0.20 & $\leq$ 0.12  \\

$^{100}$Mo  &  3.034         & $6.9 {\times} 10^{18}$   
& 8.0
& $\leq$ 0.76  & $\leq$ 0.67 & $\leq$ 0.34 & $\leq$ 0.30 
& $\leq$ 0.40  & $\leq$ 0.26 & $\leq$ 0.18 & $\leq$ 0.11  \\ \hline

\multicolumn{12}{l}{ 
$^\dagger$ Best achieved $\FWHM$ resolution at $\Qbb$ from past and
ongoing $\0nubb$ experiments~\cite{nubb-review,MJD,EXO},
} \\
\multicolumn{12}{l}{ 
~~~~~~  not including detector R\&D programs and future projects. 
} \\

\end{tabular}
\end{center}

\label{tab::2nubkg}

\end{table*}


\subsection{Limiting Irreducible Background}
\label{sect::2nubkg}


It is instructive and important to 
quantify the interplay between various
irreducible background channels to the
required exposure. 
In particular,
one such irreducible background
is the Standard Model-allowed
$\2nubb$
\begin{equation}
^N_Z \Abb ~ \rightarrow ~
_{Z+2}^{N-2}A ~ + ~ 2 e^- ~ + ~ 2 \bar{\nu}_e ~ ~.
\label{eq::2nubb}
\end{equation}
The contamination levels to $\0nubb$ at the $\Qbb$-associated RoI
depend on its half-life ($\tauhalf ^{2 \nu}$) 
and the detector resolution.
A worse resolution (larger $\Delta$) 
implies a larger RoI range to search for $\0nubb$ signals, 
and therefore a higher probability of having 
background events from the $\2nubb$ spectral tail.

Depicted in Figure~\ref{fig::Delta2nubb} are
variations of the required $\Delta$ with $\mbb$
such that $\2nubb$ background within RoI=$\FWFM$ would contribute
less than the $\BI0$-values specified by 
the benchmark and background-free conditions.
The finite width of the band 
is a consequence of the spread of
measured $\tauhalf ^{2 \nu}$~\cite{nubb-review,GERDA}.  
Faster $\2nubb$ rates
typically require better detector resolution
to define smaller RoI.
The relative locations for different $\Abb$  
within the bands are depicted in the inset.

Listed in Table~\ref{tab::2nubkg} are the
required ranges  of $\Delta$
to cover $\mbb_{95\%}^{\rm IH(NH)}$ and $\mbbmin^{\rm IH(NH)}$.
In particular,
the required resolutions  
to cover $\mbbmin$ for IH and NH
under background-free conditions
are
$\Delta  {\leq} ( 0.3 - 0.9 ) \%$
and 
$\Delta {\leq} ( 0.1 - 0.4 ) \%$, respectively.
The best achieved $\Delta$ for past and 
ongoing $\0nubb$ experiments ~\cite{nubb-review,MJD,EXO}
are included in Table~\ref{tab::2nubkg}
and depicted in the right vertical axis of
Figure~\ref{fig::Delta2nubb} for comparison.
In particular,
the best published $\Delta ( \ge76 )$=0.12\%~\cite{MJD}
corresponds to an irreducible 
$\2nubb$ background contribution of
$\BI0 {<} 6 {\times} 10^{-10} {~} \BIunit$.
This provides a comfortable margin
relative to that which satisfies the 
background-free conditions for $\mbbmin ^{\rm NH}$
at $\BI0 {\leq} 0.96 {\times} 10^{-6} {~} \BIunit$.

\section{Summary and Prospects}

As current neutrino oscillation experiments
reveal a preference of NH, 
the strategy of scaling the summit of $\0nubb$
should take this genuine possibility into account.

This work studies the relation between the
two main factors in improving experimental sensitivities:
$( \BI0 , \Sigma )$.
We recall that the presented results are derived 
with certain input parameter choice:
IA=100\%, $\varepsilon_{expt} {=} 100\%$ and
$g_A$=1.27, and that $\0nubb$ is driven by the
Majorana neutrino mass terms via the mass mechanism  
while the Signal-to-Background analysis is based on
counting experiments without exploiting
the spectral shape information at this stage.

Advancing towards ND-NH to cover $\mbbmin^{\rm NH}$
will require large and costly exposure. 
An unrealistic $\mathcal{O}$(10)~Mton-yr enriched target
mass is necessary at the current
best achieved background level $\BIcurrent {\sim} 1 {~} \Bnowunit$.
Reduction of $\BI0$ will be playing
increasingly significant, if not determining, roles 
in shaping future $\0nubb$ projects.

For instance, following Table~\ref{tab::bkgscenario},
background-free conditions for $\mbbmin^{\rm NH}$ correspond to 
additional background suppression from
the current best $\BIcurrent {\sim} 1 {~} \Bnowunit$ and 
benchmark [1~$\BM$] levels
by factors of $( 0.96 \times 10^{-6} )$ and 
$( 4.4 \times 10^{-3} )$, respectively.
This would reduce the required $\Sigma$ 
from 11~Mton-yr and 4600~ton-yr, respectively, to 550~ton-yr.
The corresponding minimal-exposure to 
cover $\mbb_{95\%}^{\rm NH}$ is 
$\Sigma_{\rm min} {\sim} 37 ~ \tyr$, 
which is only a modest factor beyond the goals of
next-generation experiments~\cite{NG0nubb}.
The pursuit of background towards 
$\BI0 {\sim} \mathcal{O} ( 10^{-6} ) {~} \BIunit$
to probe ND-NH, while challenging,
is highly investment-effective, 
as it is equivalent to reduction of $\Sigma$
by $\mathcal{O}$(10)~Mton-yr and $\mathcal{O}$(1)~kton-yr 
relative to those required for
the current best and  benchmark background levels, respectively.

This article serves to quantify the merits of background
reduction in $\0nubb$ experiments, 
but does not attempt to address the experimental
issues on how to realize the feat {\it and} how to demonstrate that
the suppression factors are achieved when experiments are constructed.
We project that the continuous intense efforts 
and ingenuities from the experimentalists world-wide,
with motivations reinforced by the increasing  
equivalent ``market'' values,
will be able to meet the challenges. 

Boosting $\Sigma$ involves mostly in the accumulation
of enriched $\Abb$ isotopes and turning these 
into operating detectors.
These processes are confined to relatively few 
locations and small communities of expertise.
The room of development which may overcome
the known hurdles is limited.
Suppression of the $\0nubb$ experimental background, 
on the other hand, would be
the tasks of mobilizing and coordinating the efforts
of a large pool of expertise. 
It is related to the advances in diverse disciplines 
from novel materials to chemistry processing to 
trace measurement techniques.
Research programs on many subjects requiring low-background 
techniques may contribute to $-$ and benefit from $-$ the
advances.  There would be strong potentials of 
technological breakthroughs and innovative ideas 
as the sensitivity goals are pursued.

Signal efficiencies are also increasingly costly
as sensitivities advance towards ND-NH.
For instance, at $\Sigma_{\rm min}$=550~ton-yr 
to cover $\mbbmin^{\rm NH}$, a high 90\% efficiency to certain
selection criterion corresponds to discarding data of
$\mathcal{O}$(10)~ton-yr strength $-$ 
already an order of magnitude
larger than the combined exposure 
of all $\0nubb$ experiments.
It follows that background suppression would preferably 
be attended at the root level $-$ that 
radioactive contaminations are suppressed to start with,
rather than relying on special signatures and
software selection algorithms to identify them. 

The next generation of $\0nubb$ experiments 
would cover $\mbbmin ^{\rm IH}$.
In addition, they should be able to
explore the strategies and demonstrate
sufficient margins to advance towards $\mbbmin ^{\rm NH}$.
A significant merit would be to have no irreducible
background before reaching the 
$\BI0 {\sim} \mathcal{O} ( 10^{-6} ) {~} \BIunit$
background-free configuration.
The detector requirements to achieve this
for $\2nubb$ are summarized in Figure~\ref{fig::Delta2nubb}
and Table~\ref{tab::2nubkg}.
Detailed studies of this background
as well as other channels  
like those due to residual cosmogenic radioactivity
and long-lived radioactive isotopes
are themes of our on-going research efforts.


\hspace*{0.5cm}

\section{Acknowledgement}

This work is supported by 
the Academia Sinica Principal Investigator Award 
AS-IA-106-M02,
contracts 104-2112-M-259-004-MY3
and 107-2119-M-001-028-MY3
from the Ministry of Science and Technology, Taiwan,
and
2017-ECP2 from 
the National Center of Theoretical Sciences, Taiwan.



\end{document}